\documentclass[apj,twocolumn, floatfix]{aastex701} \usepackage{amsmath} \usepackage{longtable} \usepackage{graphicx} \usepackage{xcolor} \usepackage{lettrine} \usepackage{yfonts} \usepackage{rotating}
\usepackage{listings}

\begin{document}

\title{\texttt{Custom Colors}: A Module for Computing Synthetic Photometry On-the-Fly in Stellar Evolution Calculations, Integrated with \texttt{MESA} and Usable with Other Stellar Evolution Codes}

\author[0000-0002-3780-0592]{Niall Miller} \affiliation{University of Wyoming, 1000 E University Ave, Laramie, WY USA} \email{niall.j.miller@gmail.com}

\author[0000-0002-8717-127X]{Meridith Joyce} \affiliation{University of Wyoming, 1000 E University Ave, Laramie, WY USA} \email{mjoyce8@uwyo.edu}

\author[0000-0001-6631-2566]{Philip Mocz} \affiliation{Center for Computational Astrophysics, Flatiron Institute, 162 Fifth Avenue New York, NY 10010, USA} \email{pmocz@flatironinstitute.org}

\begin{abstract}
Stellar evolution simulations predict physical quantities such as luminosity, whereas broadband photometric observations measure flux in specific bandpasses; converting the predicted stellar state to magnitudes requires interpolating a stellar atmosphere model at the surface parameters, convolving the resulting spectral energy distribution (SED) with filter transmission curves, and applying a photometric zero-point correction. We introduce \texttt{Custom Colors}, which performs this conversion from the stellar model. It is integrated into \texttt{MESA} as \texttt{MESA Custom Colors}, and available as the Python package \texttt{SED\_Model} for external applications. At each timestep, the module interpolates a user-specified atmosphere grid at the current $(T_{\rm eff}, \log g, [\mathrm{M/H}])$, applies geometric dilution at a user-specified distance, convolves the SED with each requested filter transmission curve, and appends observer-frame magnitudes to standard \texttt{MESA} output files. We demonstrate the module across six diverse use cases: TP-AGB evolution, nonlinear RR Lyrae pulsation, starspot-modified photospheres, white dwarf cooling, blue-loop evolution, and rotational spot modulation based on an external \texttt{YREC SPOTS} grid. These cases draw on Kurucz/ATLAS9, BT-Settl, and Koester DA atmosphere grids and filter systems including Roman WFI, LSST $ugrizy$, Gaia, and extended Johnson $UBVRIJHKLMN$. Two of these demonstrations compare the module directly with observational data: a $0.6\,M_\odot$ DA white dwarf cooling track follows the broad Gaia DR3 cooling locus from $M_G \simeq 8.5$ to $15$\,mag without intermediate color transformation, and an RSP model recovers the \textit{Kepler} light curve of FN Lyr (KIC\,6936115) to $1.1\%$ in period and $0.2\%$ in amplitude.
\end{abstract}


\section{Introduction}
\label{sec:intro}


Stellar evolution codes such as Modules for Experiments in Stellar Astrophysics \citep[\texttt{MESA};][]{Paxton2011,Paxton2013,Paxton2015,Paxton2018,Paxton2019}, the Yale Rotating Stellar Evolution Code \citep[\texttt{YREC};][]{Somers2020}, the Garching Stellar Evolution Code \citep[GARSTEC;][]{Weiss2008}, the Dartmouth Stellar Evolution Program \citep[DSEP;][]{Dotter2008}, \texttt{CESAM} \citep{Morel2008}, the PAdova and TRieste Stellar Evolution Code \citep[PARSEC;][]{Bressan2012}, BaSTI \citep{Pietrinferni2004}, and the Monash stellar evolution code \citep{Karakas2007}, among others (see \citealt{Cinquegrana2022} and references therein for a fuller overview), compute models from pre-main-sequence contraction through advanced burning and stellar endpoints, producing quantities such as luminosity, effective temperature, and surface gravity while tracing changes in composition.
Observers, by contrast, measure flux in specific bandpasses, producing magnitudes that depend on the convolution of a star's spectral energy distribution with an instrument's wavelength-dependent response.
Bridging the two typically requires interpolating a stellar atmosphere model at the surface parameters of a stellar evolutionary model, convolving the resulting spectral energy distribution (SED) with each filter transmission curve, and applying a photometric zero-point correction.
This step has traditionally been carried out offline, after the evolution calculation has finished.

Separating the photometry from the evolution introduces a number of difficulties.
The transformation is performed by stand-alone scripts that read the evolution output and depend on separately chosen atmosphere grids, interpolation schemes, and filter curves, so the same evolutionary track can yield different photometry in different hands.
For rapid phases such as thermal pulses, blue-loop crossings, or pulsation cycles, the saved model sequence must be sampled finely enough for the synthetic photometry to resolve the same physical timescales as the underlying stellar calculation.
Also, while a \texttt{MESA} \texttt{inlist} (or an equivalent user control file) specifies the physics and numerics of the evolution calculation, the subsequent photometric choices (atmosphere grid, interpolation method, filter set, and magnitude system) often live outside the place where run conditions are specified and may go undocumented, complicating reproduction and comparison between groups.

Pre-computed libraries such as \texttt{MESA} Isochrones and Stellar Tracks \citep[MIST;][]{Dotter2016,Choi2016} remain the appropriate tool for isochrone fitting and population-level comparisons, but they necessarily adopt fixed physics assumptions and a fixed photometric grid.
MSG \citep{Townsend2023} takes a complementary approach: it interpolates stellar spectra and photometric colors within a pre-calculated grid as a stand-alone post-processing step, independent of any particular evolution code.
\texttt{Custom Colors} evaluates photometry from the evolving model at runtime rather than from a saved track or a fixed library. Changes to the input physics reach the magnitudes through the surface quantities read at each step. The atmosphere grid is fixed, so this coupling runs through effective temperature, surface gravity, and surface composition alone.

Gaia has delivered parallaxes, proper motions, and multi-band photometry for nearly two billion stars \citep{gaia_edr3_2}, and the Nancy Grace Roman Space Telescope \citep{roman}, the Vera C.\ Rubin Observatory's Legacy Survey of Space and Time \citep[LSST;][]{LSST}, and the PLAnetary Transits and Oscillations of stars \citep[PLATO;][]{plato} mission will add multi-epoch photometry in a range of mission-specific filter systems.
Interpreting these data requires synthetic photometry in the native bandpasses of each survey, with temporal sampling appropriate to both the observations and the physical variability being modeled.

\vspace{0.4cm}
\noindent \texttt{Custom Colors}
\\\texttt{Custom Colors} is the photometry engine: it interpolates an atmosphere grid at the surface parameters of a stellar model, convolves the resulting SED with filter transmission curves, and applies the photometric zero-point.
The engine is available in two forms.
\texttt{MESA Custom Colors} is the implementation distributed with \texttt{MESA}, which computes photometry at runtime.
\texttt{SED\_Model} is a Python package that calls the same routines to post-process tracks from any stellar evolution code, and is described in Section~\ref{sec:sed_tools_model}.
\texttt{SED\_Tools} is a separate, complementary package that prepares atmosphere grids and filter curves into the format the engine reads.
Throughout this paper, \texttt{Custom Colors} refers to the engine and its capabilities, and \texttt{MESA Custom Colors} to the \texttt{MESA} implementation specifically.

Using \texttt{MESA Custom Colors}, at each timestep for which \texttt{MESA} writes history output, the module:
\begin{itemize}
\item[] reads the current effective temperature, surface gravity, and composition; 
\item[] interpolates a user-specified atmosphere grid to construct an SED; 
\item[] applies geometric dilution to convert surface flux to observer-frame flux at a specified distance; 
\item[] integrates the SED for the bolometric flux and magnitude; and 
\item[]convolves it with each filter transmission curve to produce magnitudes in all requested passbands.
\end{itemize}
These quantities are appended as columns in the standard \texttt{MESA} output (\texttt{history.data}) file, alongside luminosity, effective temperature, and other global or surface-averaged evolutionary quantities.

Generating the photometry during the run, rather than afterward, means that adjusting the model and re-running regenerates the magnitudes with no separate step to update, and that changes to the interior physics propagate to the magnitudes through their effect on effective temperature, surface gravity, radius, luminosity, and surface composition. 
The SEDs themselves are still drawn from external atmosphere grids, so the photometry remains limited by the coverage and assumptions of those grids.
The timestep-level alignment between structure and photometry also suits time-domain applications: asteroseismic studies that couple \texttt{MESA} structure models to GYRE pulsation calculations \citep{GYRE,2018Townsend}, for example, can evaluate mode predictions and synthetic photometry from corresponding stellar states.

The remainder of this paper is organized as follows.
Section~\ref{sec:observational} describes the observational context, Section~\ref{sec:methods} presents the machinery and user workflow, Section~\ref{sec:demonstrations} gives the demonstrations and comparisons with observations, Section~\ref{sec:limitations} discusses limitations and usage guidelines, and Section~\ref{sec:conclusions} summarizes the results and planned extensions.

\section{Observational Context}
\label{sec:observational}
Modern photometric surveys provide unprecedented precision, cadence, wavelength coverage, and sample size, placing increasing demands on the models used to interpret them.
Roman emphasizes near-infrared imaging across eight filters, targeting Galactic bulge populations, microlensing events, and resolved stellar populations in nearby galaxies with photometric precision of $\sim$0.01 mag to F184 = 26.
LSST provides the deepest wide-field time-domain coverage, imaging 18,000 deg$^2$ in six optical bands ($ugrizy$) with $\sim$3-day cadence and $\sim$800 visits per field over ten years.
PLATO prioritizes asteroseismic precision, achieving $\leq$50~ppm~h$^{-1/2}$ photometry for stars with $V\leq11$ at 25-second cadence \citep{Rauer2014}, across a single broad optical bandpass.
Table~\ref{tab:survey_characteristics} summarizes these characteristics.
Each survey targets distinct advances in stellar physics. Under the $\sim$15-minute cadence strategy modeled by \citet{Weiss2025}, Roman's Galactic Bulge Time Domain Survey is expected to deliver asteroseismic detections for hundreds of thousands of red giants, constraining ages and masses across the bulge population.
LSST's depth and near-infrared reach will expand the volume-limited sample of L and T dwarfs by roughly an order of magnitude, enabling population-level studies of the substellar luminosity function \citep{Gizis2022}.
PLATO's core program is built around asteroseismology of bright dwarfs and subgiants, delivering the stellar ages and radii needed to characterize their transiting planets.

\begin{table*}
	\centering
	\begin{tabular}{lcccc}
		\hline\hline
		Survey & Filters              & Cadence      & Depth          & Key Stellar Science$^{\text{a}}$
        \\
		\hline
		Roman  & F062--F213 (8)       & $\sim$15 min$^{\text{b}}$  & F184 $\sim$ 26 & Bulge populations                                                                       \\
		LSST   & $ugrizy$ (6)         & $\sim$3 days & $r \sim 24$ (single-visit)    & Transients, variables                                                                   \\
		PLATO  & Broad (500--1000 nm) & 25 seconds   & $V \sim 11$    & Asteroseismology                                                                        \\
		\hline
	\end{tabular}
    \caption{Filter coverage, representative cadence and depth, and examples of key stellar-science goals for Roman, LSST, and PLATO. $^{\text{a}}$Examples of key science, not necessarily the only scientific goal. $^{\text{b}}$Cadence strategy modeled by \citet{Weiss2025}.\label{tab:survey_characteristics}}
\end{table*}

\texttt{Custom Colors} provides a form of pathfinding in which theoretical predictions can guide target selection and observing strategy before observations commence.

Multi-band light curves encode chromatic information that discriminates between physical mechanisms.
For example, pulsations produce wavelength-dependent phase lags, starspots cause stronger modulation in blue bands than in red bands, and transient outbursts trace temperature evolution through color changes.
Synthetic light curves must therefore capture this chromatic structure by convolving time-dependent SEDs with each filter's transmission curve.
The demonstrations in Section~\ref{sec:demonstrations}, summarized in Table~\ref{tab:demo_summary}, are selected to exercise these capabilities.

\section{\texttt{Custom Colors} Machinery}
\label{sec:methods}

\texttt{MESA Custom Colors} is distributed with \texttt{MESA} release \texttt{r26.04.1}\footnote{\url{https://github.com/MESAHub/mesa/releases/tag/26.4.1}}, the version used in this work.

\subsection{Computational Pipeline}
\label{sec:pipeline}
\texttt{MESA Custom Colors} computes synthetic photometry at runtime by coupling the stellar structure output with pre-processed atmosphere grids and filter transmission curves.
The synthetic photometry calculation occurs at each \texttt{MESA} timestep for which history output is written.
The pipeline begins by extracting the star's current surface parameters from the \texttt{MESA} evolution state: effective temperature ($T_{\rm eff}$), surface gravity ($\log g$), stellar radius ($R_*$), and the photospheric metal-to-hydrogen ratio, $(Z/X)_{\rm surf}$, which is converted to [M/H] using the user-selected reference ratio \texttt{z\_over\_x\_ref}.
These parameters are then used to query a user-specified library of stellar atmosphere models.
The module performs Hermite interpolation\footnote{Other available interpolation methods include linear interpolation and k-nearest neighbors.} within this pre-computed grid across the three-dimensional parameter space of ($T_{\rm eff}$, $\log g$, [M/H]) to construct a specific spectral energy distribution (SED) representing the star's surface flux as a function of wavelength, $F_\lambda(\lambda)$.
The Euclidean distance to the nearest SED (\texttt{Interp\_rad}) is recorded as a grid-proximity diagnostic.

This interpolation uses either tabulated atmosphere spectra stored as individual text files or a pre-computed flux cube stored in binary format.
If a valid \texttt{flux\_cube.bin} is present, the module attempts to allocate and load the complete four-dimensional flux array at initialization. If the file is missing or malformed, or if the allocation fails, the module reports the reason -- including the requested memory for an allocation failure -- and automatically falls back to slower per-file SED loading.
Once the surface SED is constructed, geometric dilution is applied to convert surface flux to observer-frame flux.
The dilution factor accounts for both the stellar radius and the distance to the observer: $F_{\rm obs}(\lambda) = F_\lambda(\lambda) \times (R_*/d)^2$, where $d$ is the user-specified distance (defaulting to 10 parsecs for absolute magnitudes).
The diluted SED is then integrated over all wavelengths to compute the bolometric flux, $F_{\rm bol} = \int F_{\rm obs}(\lambda) \, d\lambda$, from which the bolometric magnitude is derived as $M_{\rm bol} = -2.5 \log_{10}(F_{\rm bol}/f_0)$, where $f_0 = 2.518\times10^{-5}\,\mathrm{erg\,s^{-1}\,cm^{-2}}$ is the IAU bolometric zero-point flux. At the default 10~pc this reproduces $M_{\rm bol} = 4.74 - 2.5\log_{10}(L/L_\odot)$.
For bandpass photometry, the module convolves the observer-frame SED with each filter transmission curve specified in the user-selected instrument directory.
Each filter file defines a wavelength-dependent throughput function $S_X(\lambda)$, representing the combined efficiency of telescope optics, filter glass, and detector quantum efficiency.

The synthetic flux in filter $X$ is computed as:
\begin{equation}
	F_X = \frac{\int F_{\rm obs}(\lambda) S_X(\lambda) \, \lambda \, d\lambda}{\int S_X(\lambda) \, \lambda \, d\lambda}.
\end{equation}
This flux is converted to a magnitude using the user-selected photometric system (AB, ST, or Vega).

For the Vega system, the zero-point is defined such that the star Vega has magnitude zero in all bands:
\begin{equation}
	m_X = -2.5 \log_{10}\left(\frac{\int F_{\rm obs}(\lambda) S_X(\lambda) \, \lambda \, d\lambda}{\int F_{\rm Vega}(\lambda) S_X(\lambda) \, \lambda \, d\lambda}\right),
\end{equation}
where $F_{\rm Vega}(\lambda)$ is the reference Vega spectrum provided by the user.

For AB magnitudes, the zero-point is calibrated to a flat spectrum with $F_\nu = 3631$ Jy, while ST magnitudes use a flat spectrum with constant $F_\lambda$.
The module appends three fixed columns to the \texttt{MESA} \texttt{history.data} file--\texttt{Mag\_bol}, \texttt{Flux\_bol}, and \texttt{Interp\_rad}--along with one synthetic magnitude column for each filter discovered in the instrument directory.
The \texttt{Interp\_rad} diagnostic reports the Euclidean distance between the requested stellar parameters and the nearest tabulated atmosphere after each active grid coordinate is linearly rescaled to the unit interval spanned by that coordinate in the adopted atmosphere library. 
It is therefore a grid-dependent proximity flag, rather than a calibrated photometric uncertainty or a physically scaled distance in ($T_{\rm eff}$, $\log g$, [M/H]) space.

\subsection{Stellar Atmosphere Models and Interpolation}
\label{sec:atmosphere_grids}
The fidelity of the synthetic photometry depends critically on the stellar atmosphere grid used to construct the SEDs.
\texttt{Custom Colors} requires that atmosphere grids be organized in a standardized directory structure containing three essential components: a \texttt{lookup\_table.csv} file mapping individual atmosphere filenames to their physical parameters ($T_{\rm eff}$, $\log g$, [M/H]),
a set of SED files providing wavelength-flux pairs for each grid point, and optionally a \texttt{flux\_cube.bin} binary file encoding the entire grid in a format optimized for rapid interpolation.
The optional \texttt{flux\_cube.bin} stores the temperature, gravity, metallicity, and wavelength axes together with the corresponding dense four-dimensional flux array in a single unformatted binary stream. Loading this array once at initialization avoids repeated text parsing and disk access, while its regular in-memory layout permits vectorized interpolation across wavelength.
The atmosphere grid defines the physical regime over which the module can reliably compute photometry.
Grid coverage varies by model family; for example, the \citet{Kurucz1970,Castelli04} grid spans temperatures from 3,500 K to 50,000 K, surface gravities from $\log g = 0.0$ to 5.0, and metallicities from [M/H] = $-$5.0 to +1.0 over its full published extent.
Users working with cooler stars, such as M dwarfs or AGB stars, may require atmosphere models that extend to lower temperatures and include molecular opacity sources such as TiO, H$_2$O, and CN.
When a stellar model's parameters fall within the grid boundaries, the module performs Hermite interpolation to construct an SED at the exact ($T_{\rm eff}$, $\log g$, [M/H]) requested.
If the model falls outside the grid, the module clamps to the nearest edge point and reports this via the \texttt{Interp\_rad} diagnostic.
Users should exercise caution when interpreting photometry computed under significant clamping (large \texttt{Interp\_rad}), as atmosphere physics may differ substantially outside the regime where models were validated.

\subsection{Filter Transmission Curves and Photometric Systems}
\label{sec:filters}

Filter transmission curves encode the wavelength-dependent response of an observational system, combining the effects of mirror reflectivity, filter glass absorption, and detector quantum efficiency into a single dimensionless throughput function.
\texttt{Custom Colors} expects filter files to be plain-text, two-column data files (wavelength in Angstroms, transmission between 0 and 1) with a single header line.
Filter sets are organized hierarchically as \texttt{facility/instrument} directories; for example, the Gaia filter set resides in \texttt{GAIA/GAIA} and contains transmission curves for the G, G$_{\rm BP}$, G$_{\rm RP}$, and G$_{\rm RVS}$ bands (JWST's MIRI resides in \texttt{JWST/MIRI}).
The module discovers filters dynamically by reading an index file located within the instrument directory.
This index file (named identically to the instrument directory) lists one filter filename per line.
Each filter filename (excluding the \texttt{.dat} extension) becomes a column name in the output \texttt{history.data} file, enabling users to add or remove filters by simply editing the index file and updating the filter directory contents.
Magnitude zero-points are handled through the \texttt{mag\_system} parameter, which accepts three conventions: Vega, AB, and ST.
The Vega system defines zero magnitude as the flux from the star Vega in each band, requiring the user to provide a reference Vega spectrum via the \texttt{vega\_sed} parameter.

\subsection{Grid Preparation and the Stand-alone Python Interface}
\label{sec:sed_tools_model}
\texttt{Custom Colors} relies on pre-processed atmosphere grids and filter transmission curves supplied in the formats described above.
In this work, these input data products were prepared using \texttt{SED\_Tools}\footnote{\url{https://github.com/nialljmiller/SED_Tools}}, a Python package developed alongside \texttt{Custom Colors}.
For the applications presented here, \texttt{SED\_Tools} was used offline to standardize atmosphere spectra, assemble the corresponding \texttt{lookup\_table.csv} files, construct the pre-computed \texttt{flux\_cube.bin} files used for interpolation, and organize filter transmission curves in the directory structure expected by the module.
The Kurucz/ATLAS9 \citep{Kurucz1970,Castelli04}, BT-Settl \citep{Allard2014}, and Koester DA \citep{Koester2010} grids used in this work were all obtained from the SVO Theoretical Spectra Service and standardized by \texttt{SED\_Tools} onto common physical units and a common wavelength sampling before being packaged into the format \texttt{Custom Colors} expects.
The package is part of the data-preparation workflow rather than a runtime dependency of \texttt{MESA}. Once a grid directory and filter set have been prepared, they are selected in the \texttt{MESA} \texttt{inlist} through the \texttt{stellar\_atm} and \texttt{instrument} controls.

\texttt{SED\_Model}\footnote{\url{https://github.com/nialljmiller/SED_Model}} is the second form in which the engine is distributed.
It calls the same Fortran routines as \texttt{MESA Custom Colors} through a Python and CLI interface, and reads the same atmosphere grids and filter definitions under the same magnitude-system conventions.
A track from any stellar evolution code can therefore be passed through \texttt{Custom Colors} without an active \texttt{MESA} run, provided it supplies effective temperature, surface gravity, and composition at each step.
Section~\ref{sec:demo_yrec} applies this to a \texttt{YREC} grid.

\subsection{User Considerations}
\label{sec:user_workflow}

\texttt{MESA Custom Colors} is configured through a dedicated \texttt{\&colors} namelist block in the \texttt{MESA} \texttt{inlist}.
This module-level block sits alongside namelists such as \texttt{\&kap} and \texttt{\&eos} and is separate from the standard \texttt{\&controls} and \texttt{\&star\_job} namelists.
The master on/off switch is \texttt{use\_colors}, which when set to \texttt{.true.} activates all synthetic photometry calculations.
Users specify the atmosphere grid directory via \texttt{stellar\_atm} and the filter set via \texttt{instrument}.
The \texttt{distance} parameter sets the source distance in centimeters, defaulting to $3.0857 \times 10^{19}$ cm (10 parsecs) to produce absolute magnitudes; users modeling specific systems can set this to the actual source distance to compute apparent magnitudes.
The \texttt{mag\_system} parameter selects the photometric zero-point convention (AB, ST, or Vega), and if using Vega, the \texttt{vega\_sed} parameter must point to a reference Vega spectrum file (Appendix~\ref{app:spectrum_format}).
Optional diagnostic outputs can be enabled via \texttt{make\_csv = .true.}, which writes the full SED and filter-convolved flux to CSV files in the directory specified by \texttt{colors\_results\_directory} (where \texttt{sed\_per\_model} controls whether the same SED is repeatedly overwritten or, if \texttt{.true.}, the model number is appended to the end of the file name, producing one CSV file per model per filter).
Photometry columns are appended to \texttt{history.data} whenever \texttt{MESA} writes history output.
The sampling cadence of the recorded photometry is therefore controlled by the standard \texttt{MESA} parameter \texttt{history\_interval}; this controls the output cadence, not the temporal resolution of the underlying evolution calculation.
Users requiring high-cadence photometric time series (e.g., for pulsating stars or rapid evolutionary phases) should set \texttt{history\_interval = 1} to compute photometry at every timestep, though this increases computational overhead and file sizes (Appendix~\ref{app:timing}).
Detailed configuration examples and complete \texttt{inlist} templates are provided in Appendix~\ref{app:config}.

\section{Demonstrations}
\label{sec:demonstrations}

\begin{table*}
	\centering
	\footnotesize
	\begin{tabular}{lllll}
		\hline\hline
		Demo & Physical regime & Atmosphere grid & Filter system & Mag.\ system \\
		\hline
		TP-AGB
		     & He-shell flash cycles
		     & BT-Settl
		     & Johnson $UBVRIJHKLMN$
		     & Vega
          \\
		\\
		RR~Lyr (RSP)
		     & Nonlinear radial pulsation
		     & Kurucz/ATLAS9
		     & Roman WFI; Kepler
		     & AB
		     \\
		\\
		Starspots
		     & Two-temperature photosphere
		     & BT-Settl
		     & LSST $ugrizy$
		     & AB
          \\
		\\
		White dwarf
		     & Extreme $\log g$, degenerate EOS
		     & Koester DA
		     & Gaia
		     & Vega
		     \\
		\\
		Blue loops
		     & Wide $T_{\rm eff}$ excursion
		     & Kurucz/ATLAS9
		     & LSST $ugrizy$
		     & AB
		     \\
		\\
		YREC SPOTS
		     & Rotational spot modulation
		     & Kurucz/ATLAS9
		     & Johnson $UBVRIJ$
		     & AB
          \\
		\hline
	\end{tabular}

	\caption{Summary of the \texttt{Custom Colors} demonstrations in Section~\ref{sec:demonstrations}.
		The first five use the module at runtime inside \texttt{MESA}; the \texttt{YREC SPOTS} test (Section~\ref{sec:demo_yrec}) uses the stand-alone \texttt{SED\_Model} interface as a post-processor for a grid \texttt{MESA} did not produce.}
	\label{tab:demo_summary}
\end{table*}

The demonstrations that follow target the filter systems of the surveys shown in Table~\ref{tab:survey_characteristics}, producing forward models directly comparable to survey data.
The six demonstrations are chosen to exercise the capabilities set out in Section~\ref{sec:observational} and to represent the broad range of observable scenarios in stellar physics that \texttt{MESA} can model.
Three of the runtime tests (TP-AGB, starspots, and blue loops) are \emph{pathfinding} demonstrations that forward-model rapid, chromatic, or large-amplitude phases in the relevant survey bands ahead of the data.
We also include validations against real data: the RR~Lyrae test fits the \textit{Kepler} light curve of FN~Lyr (Section~\ref{sec:rsp_fnlyr}), and the white dwarf test places a \texttt{MESA} cooling track on the observed Gaia cooling locus (Section~\ref{sec:demo_wd}).
The \texttt{YREC SPOTS} test (Section~\ref{sec:demo_yrec}) exercises \emph{custom-physics exploration}: it applies the same photometry engine, through \texttt{SED\_Model}, to an external grid that \texttt{MESA} never computed.

The filter choice in each test is set by the physics being investigated and is stated in the corresponding section: the RR~Lyrae demonstration uses Roman WFI \citep{roman}, whose F062--F213 baseline samples the progression from near the SED peak toward the Rayleigh--Jeans regime at RR~Lyrae temperatures, plus the \textit{Kepler} band for the FN~Lyr fit; the starspot and blue-loop demonstrations use LSST $ugrizy$ \citep{LSST}; the TP-AGB demonstration uses extended Johnson $UBVRIJHKLMN$, matched to AGB photospheric temperatures; and the white dwarf validation uses the Gaia bands for the observational comparison.

\subsection{Thermally Pulsing AGB: Rapid Photometric Evolution}
\label{sec:demo_tp_agb}

Thermal pulses on the asymptotic giant branch are periodic thermonuclear runaways in the helium-burning shells of evolved, low- to intermediate-mass stars. They are of particular interest in \texttt{Custom Colors} applications because thermal pulses are among the few stellar evolutionary events that can be observed on human timescales, as illustrated by the observed period evolution of T~Ursae Minoris and R~Hydrae \citep{Molnar2019,Joyce2024}.
Each flash drives a surface luminosity increase of a factor $\sim$2--10 over $\sim$100--1000~years, embedded in quiescent interpulse phases of $\sim$50{,}000--100{,}000~years, accompanied by radius expansion and a corresponding photometric excursion of $\Delta V \sim 0.5$--1.5~mag.

The challenge for synthetic photometry here is that the flash rise is short compared with the interpulse evolution, so the saved model sequence must be sampled finely enough to follow the corresponding photometric response.
In this demonstration, we use runtime photometry with \texttt{history\_interval = 1}, so the magnitudes are written from the same saved stellar states used to describe the pulse evolution.

The flash also stresses the convective treatment.
When the flash timescale falls below the convective turnover time in the intershell region, convection cannot adjust instantaneously, and \texttt{MESA}'s time-dependent convection \citep[TDC;][]{Paxton2011,Paxton2013,Paxton2015,Paxton2018,Paxton2019} evolves the convective velocity as a dynamical variable rather than assuming equilibrium mixing-length theory (MLT).
Because the surface temperature and radius depend on the convective response during the flash, runtime photometry records whatever convective state the structure model produces, with no separate assumption imposed at the post-processing stage.

We evolve a $2\,M_\odot$, solar-metallicity model from the pre-main sequence through core hydrogen and helium burning to the thermally pulsing AGB.
The first stage (\texttt{MESA Custom Colors} disabled) establishes the structure at the onset of pulses; the second continues with \texttt{use\_colors = .true.} and \texttt{history\_interval = 1}, recording photometry at every timestep.
We adopt extended Johnson $UBVRIJHKLMN$ with BT-Settl atmospheres \citep{Allard2014}: AGB photospheric temperatures reach $\sim$3300--3500~K at interpulse minima and cooler at pulse maximum, below the $\sim$3500~K floor where Kurucz models become unreliable, whereas BT-Settl remains valid across this range with the appropriate molecular opacities.
Timestep controls (\texttt{delta\_lgL\_limit = 0.05}, \texttt{delta\_lgTeff\_limit = 0.01}) finely sample the rapid phases, yielding hundreds of photometric points per flash.

\begin{figure}[tbp]
	\centering
	\includegraphics[width=\columnwidth]{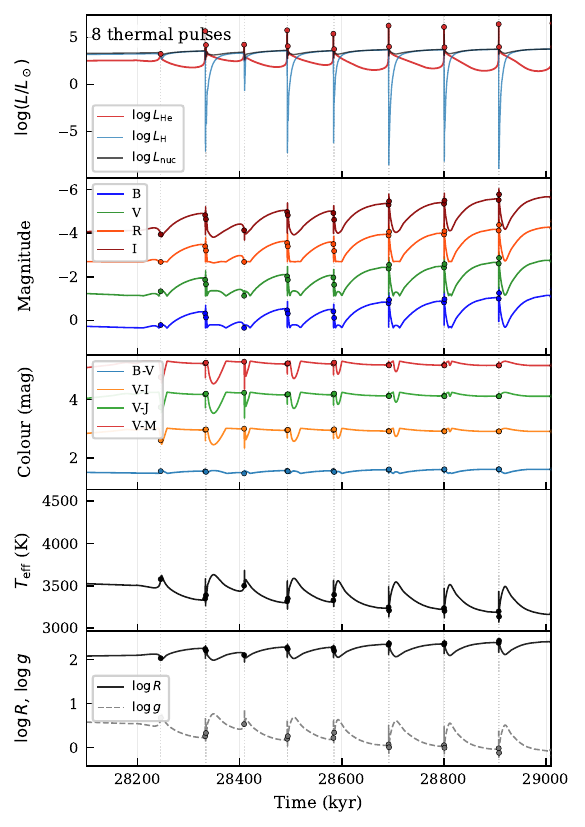}
	\caption{Synthetic photometry across eight thermal-pulse cycles for the $2\,M_\odot$ TP-AGB model.
		Vertical dotted lines mark pulse maxima.
		Top to bottom: (1)~internal luminosity components ($\log L_{\rm He}$, $\log L_{\rm H}$, $\log L_{\rm nuc}$); (2)~Johnson $BVRI$ magnitudes; (3)~color indices ($B-V$, $V-I$, $V-J$, $V-M$); (4)~$T_{\rm eff}$; (5)~$\log R$ (solid) and $\log g$ (dashed).
        The high-cadence sampling (\texttt{history\_interval = 1}) records the photometric response throughout the rapid rise of each pulse at the same cadence as the \texttt{MESA} history output.}
	\label{fig:tp_agb_lightcurves}
\end{figure}

Figure~\ref{fig:tp_agb_lightcurves} shows the photometric evolution across eight pulse cycles.
The internal luminosity components (top panel) connect the deep instability to the surface response: $\log L_{\rm He}$ spikes by orders of magnitude at each flash onset, while $\log L_{\rm H}$ and $\log L_{\rm nuc}$ respond on longer timescales.
The Johnson magnitudes (second panel) brighten by $0.5$--$1.5$~mag at each pulse and recover slowly, with the amplitude systematically larger in bluer bands, as expected from the stronger Wien-regime temperature sensitivity at this photosphere.
The color indices (third panel) track the reddening toward each maximum and the recovery as the envelope contracts.
The fourth panel is an internal-consistency check: the bolometric magnitude integrated from the \texttt{MESA Custom Colors} SED agrees at all phases with $M_{\rm bol} = 4.74 - 2.5\log(L/L_\odot)$ derived from the structure $L$.
The lower panels give $T_{\rm eff}$ and the structural quantities ($\log R$, $\log g$).
Over the full cycle $T_{\rm eff}$ varies by several hundred kelvin (panel 5), the upper end of the $\Delta T_{\rm eff}$ range quoted above.

\begin{figure*}[t]
	\centering
	\includegraphics[width=\textwidth]{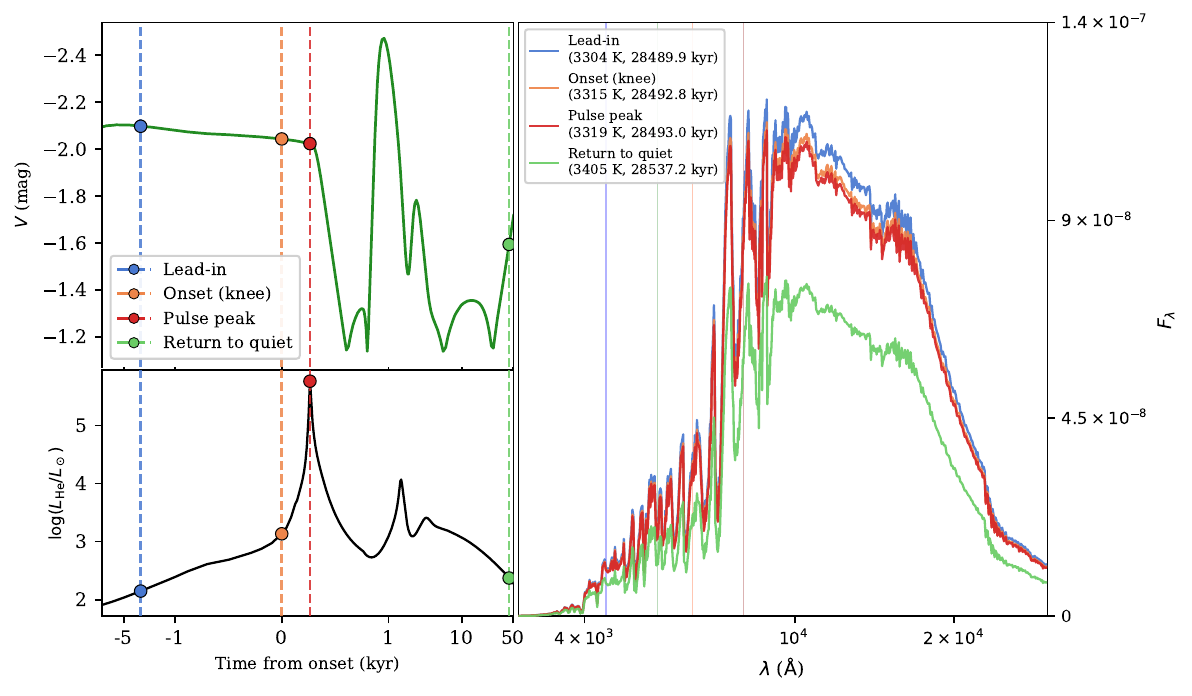}
	\caption{Phase-resolved view of a single thermal pulse for the $2\,M_\odot$ TP-AGB model.
	\textit{Left}: $V$-band magnitude (top) and $\log(L_{\rm He}/L_\odot)$ (bottom) versus time relative to onset, on a symmetric-log axis.
	Four phases are marked consistently: Lead-in, Onset (knee), Pulse peak, Return to quiet.
	\textit{Right}: BT-Settl SEDs at the four phases over 3000--30{,}000~\AA, with Johnson $BVRI$ locations as faint vertical lines; legend entries give phase, $T_{\rm eff}$, and model time.
	The four sampled phases span only $\Delta T_{\rm eff} \approx 100$~K (the wider full-cycle swing falls between them, on the interpulse recovery); the pulse-peak brightening is dominated by radius expansion, not temperature.}
	\label{fig:tp_agb_single_pulse}
\end{figure*}

Figure~\ref{fig:tp_agb_single_pulse} resolves one pulse in detail. It shows how the helium-shell diagnostic compares to the $V$-band response and the underlying SED.
The left column shows $V$-band magnitude and $\log(L_{\rm He}/L_\odot)$ on a symmetric-log time axis that holds both the $\sim$1~kyr rise and the $\sim$50~kyr recovery in one frame.
Four phases are marked consistently across panels: Lead-in, Onset (knee), Pulse peak (helium luminosity maximum), and Return (decay) to quiet.
The right panel overlays the emergent BT-Settl SEDs at these four phases.
The phase-to-phase SED differences are concentrated at short wavelengths, making $B$ the most responsive band during the flash.
The Return-to-quiet SED is the warmest and bluest, reflecting post-flash contraction.

The flash timescale is resolved by \texttt{MESA}'s adaptive timestepping, so the light curve carries no temporal interpolation, and the TDC convective response is recorded directly in the magnitudes at each step.

\subsection{Radial Stellar Pulsations: RR Lyrae}
\label{sec:demo_rsp}

Classical pulsating variables are a stringent test of synthetic photometry in a time-dependent regime: the stellar state changes at every timestep, and the emergent magnitudes must respond correctly at each one.
We use RR~Lyrae pulsation to validate \texttt{MESA Custom Colors} in two stages.
First, in a controlled model where $T_{\rm eff}$ and $R$ are known at every timestep, we verify that \texttt{MESA Custom Colors} recovers the known wavelength dependence of RR~Lyrae observables from a nonlinear RSP calculation (Section~\ref{sec:rsp_roman}).
Second, we fit an RSP model, through its \texttt{MESA Custom Colors} light curve, to the \textit{Kepler} photometry of a real RR~Lyrae star (Section~\ref{sec:rsp_fnlyr}).

\texttt{MESA}'s RSP module \citep[RSP;][]{Smolec2008} computes nonlinear radial pulsation with a time-dependent convective treatment, producing self-consistent variations in $T_{\rm eff}$, $L$, and $R$ at each hydrodynamic timestep.
With \texttt{MESA Custom Colors} enabled, synthetic photometry is generated at the native cadence of the pulsation calculation, with no temporal interpolation and exact alignment between the stellar state and the emergent magnitudes.

\subsubsection{Recovering the Wavelength Dependence: Roman WFI}
\label{sec:rsp_roman}
The first test demonstrates the wavelength dependence of RR~Lyrae pulsation that follows from the temperature sensitivity of the Planck function: the decline of pulsation amplitude from the optical to the near-infrared, and the distinct ways the broadband color tracks effective temperature and radius.
We use the Roman Wide Field Instrument's imaging filters (F062--F213), spanning 0.48--2.30~$\mu$m, from the near-optical regime close to the SED peak of a $\sim$6500~K star, where flux is strongly temperature-sensitive, toward the near-infrared Rayleigh--Jeans regime, where the sensitivity is much weaker.
A single mission filter set therefore samples the broad progression from near the SED peak toward Rayleigh--Jeans behavior at RR~Lyrae temperatures.

Roman's Galactic Bulge Time Domain Survey will deliver multi-band light curves for the dense bulge RR~Lyrae population \citep{roman}, so this is the behavior that survey will need to interpret.
Because the aim is to check the synthetic photometry against a known temperature, we adopt a representative fundamental-mode model rather than a fit to a particular star.

We compute a fundamental-mode RR~Lyrae model with $M = 0.65\,M_\odot$, $L = 50\,L_\odot$, $T_{\rm eff} = 6600$~K, and $Z = 0.0004$.
The model is first relaxed to a stable limit cycle; \texttt{MESA Custom Colors} is then enabled (\texttt{RSP\_target\_steps\_per\_cycle = 100}, \texttt{RSP\_max\_dt = 600\,s}), using the Roman WFI filters, the Kurucz/ATLAS9 grid \citep{Castelli04}, and the AB system.
On the settled limit cycle the period is $P = 0.572$~d and the growth diagnostic is $|\mathrm{rsp\_GREKM}| \approx 4\times10^{-5}$, confirming a stable cycle. Over a cycle the model spans $T_{\rm eff} = 6136$--$7514$~K ($\Delta T_{\rm eff} = 1378$~K), median $\log(L/L_\odot) = 1.63$, and a photospheric radius of $4.99$--$5.78\,R_\odot$.
The surface becomes supersonic on the rapid rise, as expected for a large-amplitude RRab pulsation; we report the photometric behavior of this self-consistent model and do not tune its morphology to any individual star.
All quantities below are measured on the settled cycles, after the kick-driven transient has decayed.

\begin{figure}[tbp]
	\centering
	\includegraphics[width=\columnwidth]{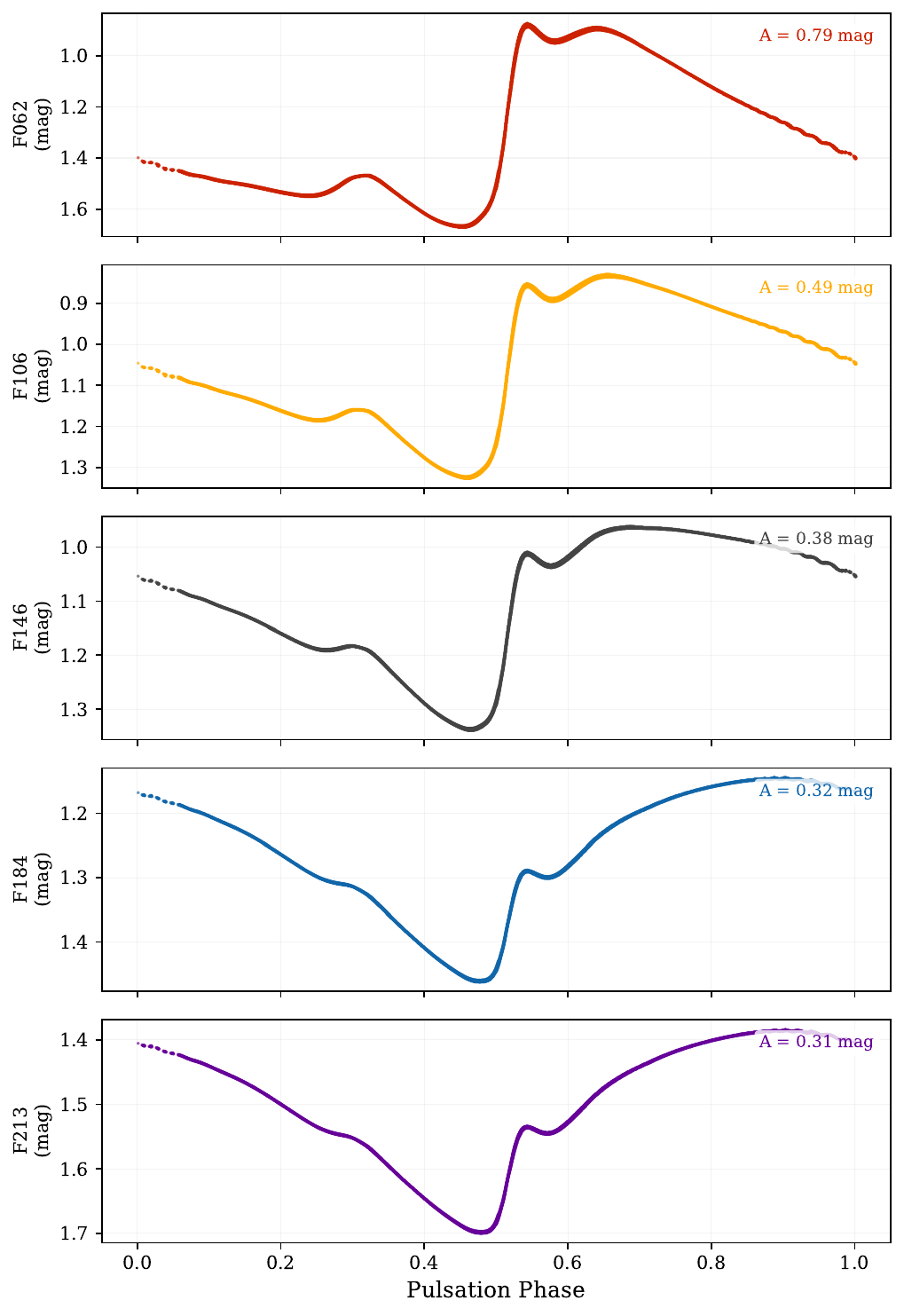}
	\caption{Phase-folded Roman WFI light curves for the settled RR~Lyrae model, in F062, F106, F146, F184, and F213 (top to bottom).
		The peak-to-peak amplitude, annotated in each panel, decreases monotonically from $0.79$~mag in F062 to $0.31$~mag in F213, and the phase of maximum light shifts redward, reflecting the transition from temperature-dominated to increasingly radius-influenced response across the near-optical to near-infrared baseline.}
	\label{fig:rsp_lightcurves}
\end{figure}

The clearest signature of the wavelength dependence is the run of pulsation amplitude across the Roman bands.
Figure~\ref{fig:rsp_lightcurves} shows the phase-folded light curves in F062, F106, F146, F184, and F213.
The peak-to-peak amplitude decreases monotonically toward the red, from $0.79$~mag in F062 to $0.49$~mag in F106, $0.38$~mag in F146, $0.32$~mag in F184, and $0.31$~mag in F213, with the remaining bands ordered by wavelength between these values.
This follows directly from the Planck function's steeper temperature sensitivity at shorter wavelengths: the same $T_{\rm eff}$ swing drives a larger fractional flux change near the SED peak, sampled by F062, than toward the less temperature-sensitive Rayleigh--Jeans regime sampled by F213.
The phase of maximum light also migrates redward, from $\phi \approx 0.54$ in F062 to $\phi \approx 0.90$ in F213, near the phase of maximum radius.
The bluest bands therefore peak with temperature while the reddest carry a growing radius contribution, so the phase of maximum light is itself a wavelength-dependent diagnostic of the pulsation.

\begin{figure*}[t]
	\centering
	\includegraphics[width=\linewidth]{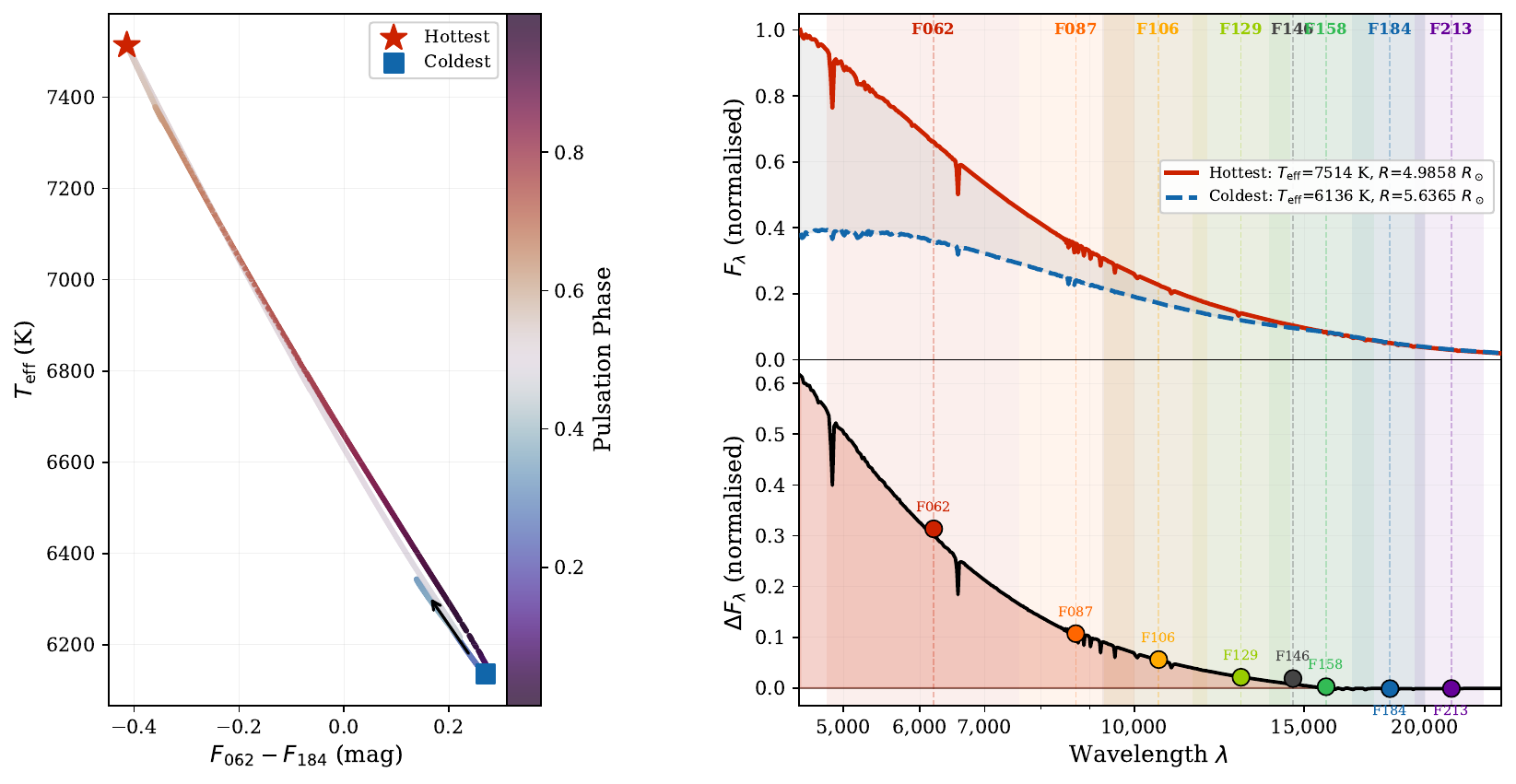}
	\caption{Temperature diagnostic in Roman WFI, from the \texttt{MESA Custom Colors} SEDs.
		\textit{Left}: $T_{\rm eff}$ versus $(F_{062} - F_{184})$ over the settled cycle, phase-colored; the near-linear, single-valued relation ($r = -0.999$) shows the color is a reliable temperature proxy at every phase, with the hottest ($7514$~K) and coldest ($6136$~K) moments marked.
		\textit{Upper right}: interpolated SEDs at these moments, normalized to the hotter peak, with Roman bandpasses shaded.
		\textit{Lower right}: residual $\Delta F_\lambda = F_{\rm hot} - F_{\rm cold}$, decreasing monotonically from F062 to F213 (filter-integrated values as circles).}
	\label{fig:rsp_teff_sed}
\end{figure*}

The broadband color $(F_{062} - F_{184})$ combines the most and least temperature-sensitive Roman bands and should track $T_{\rm eff}$ closely.
F062 lies near the SED peak of a $\sim$6500~K star, where the Planck function's logarithmic temperature derivative $\partial \ln B_\lambda / \partial \ln T = x e^x/(e^x-1)$ (with $x \equiv hc/\lambda kT$) is large, whereas toward F184 the derivative is substantially smaller, though still above its Rayleigh--Jeans limit of unity; a hotter star therefore brightens proportionally more at F062 than at F184, driving the color bluer.
Figure~\ref{fig:rsp_teff_sed} (left) shows $T_{\rm eff}$ against this color over the settled cycle: the relation is near-linear and single-valued, and the Pearson correlation between the un-normalized quantities is $r = -0.999$.
The right panels show the mechanism using the interpolated SEDs written to disk during the run, not a blackbody: the hotter SED ($T_{\rm eff} = 7514$~K) lies above the cooler one ($6136$~K) at all wavelengths, but their separation is largest in the blue and converges toward zero in the infrared, so the residual $\Delta F_\lambda = F_{\rm hot} - F_{\rm cold}$ decreases monotonically from F062 to F213.
The filter-integrated values reproduce this ordering across all eight bands.

\begin{figure*}[t]
	\centering
	\includegraphics[width=\linewidth]{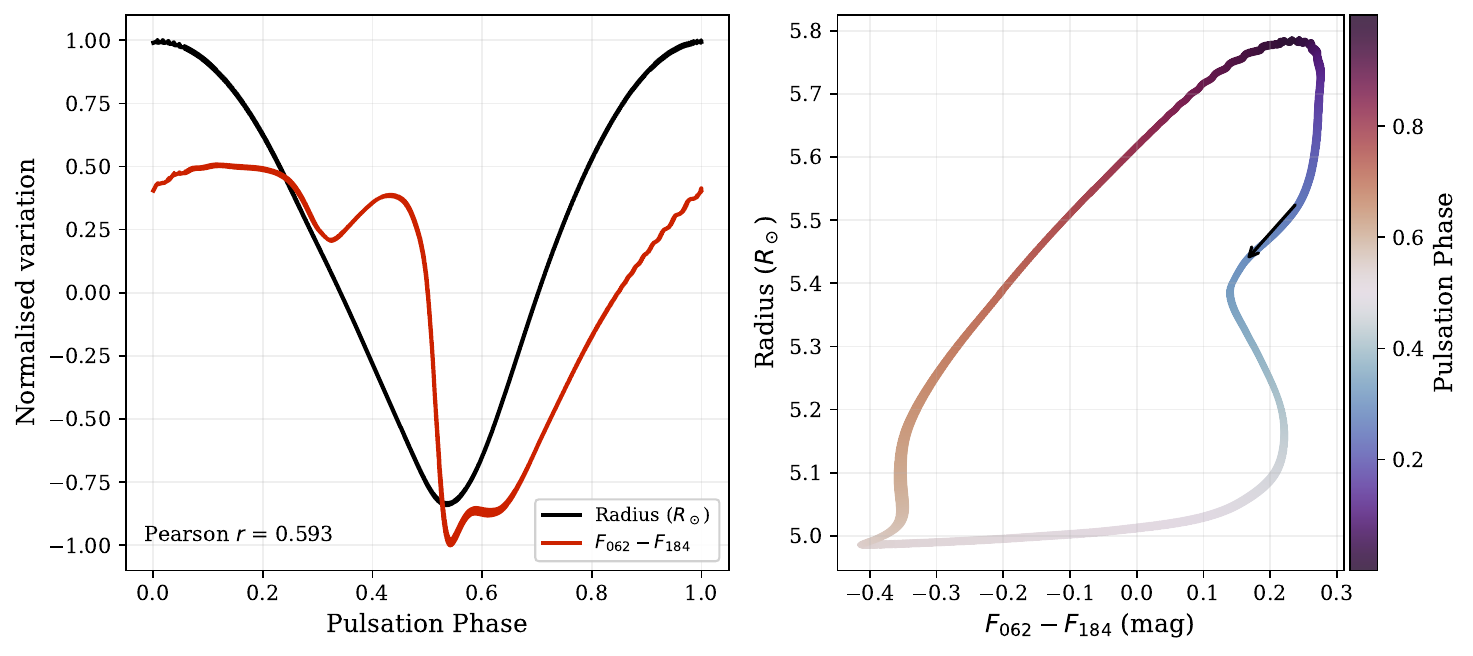}
	\caption{Radius diagnostic in Roman WFI for the settled cycle.
		\textit{Left}: normalized radius (black) and normalized $(F_{062} - F_{184})$ (red) versus phase; the two are only moderately correlated ($r = 0.59$), against $-0.999$ for the temperature--color pair (Figure~\ref{fig:rsp_teff_sed}).
		\textit{Right}: radius versus color, phase-colored; the open loop reflects the temperature--radius phase offset.}
	\label{fig:rsp_radius_diagnostic}
\end{figure*}

Because temperature and radius vary out of phase over the cycle, the color and radius are only moderately correlated and trace a hysteresis loop rather than a single-valued relation.
A temperature proxy should not, in turn, track radius directly.
Figure~\ref{fig:rsp_radius_diagnostic} shows this: the un-normalized Pearson correlation between color and radius is $r = 0.59$, against $-0.999$ for color and temperature, and the right panel traces a broad open loop with distinct branches at the same color.
At a given color the radius depends on cycle phase.

\begin{figure*}[t]
	\centering
	\includegraphics[width=\linewidth]{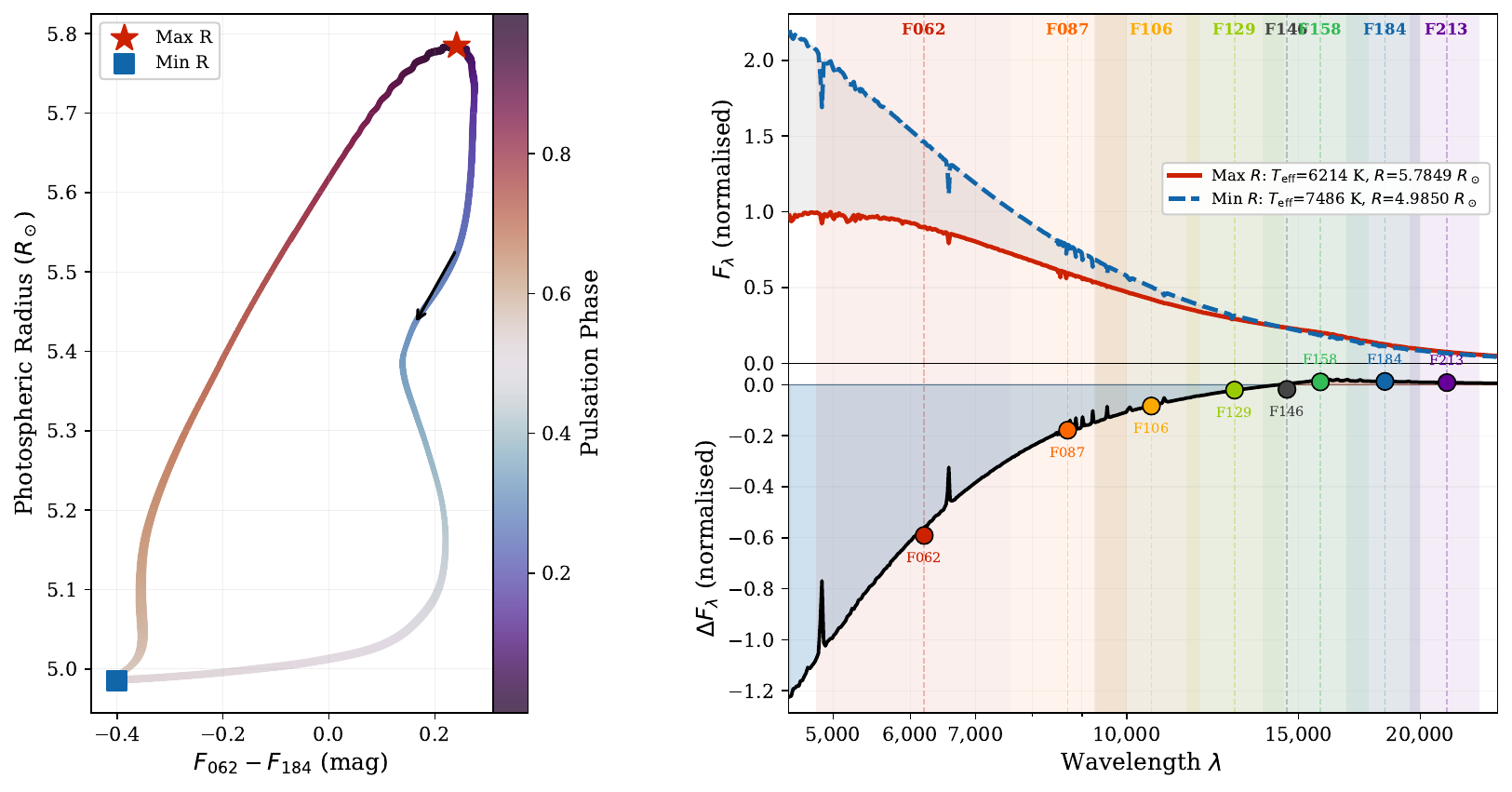}
	\caption{Radius variation and spectral residual in Roman WFI, from the \texttt{MESA Custom Colors} SEDs.
		\textit{Left}: radius versus $(F_{062} - F_{184})$ over the settled cycle, phase-colored, with the maximum-radius ($T_{\rm eff} = 6214$~K, $R = 5.78\,R_\odot$) and minimum-radius ($T_{\rm eff} = 7486$~K, $R = 4.99\,R_\odot$) moments marked.
		\textit{Upper right}: interpolated SEDs at these moments.
		\textit{Lower right}: residual $\Delta F_\lambda = F_{\rm max\,R} - F_{\rm min\,R}$, strongly negative in the blue and crossing zero near F158 as the larger radius offsets the lower temperature.}
	\label{fig:rsp_radius_sed}
\end{figure*}

Figure~\ref{fig:rsp_radius_sed} examines the SEDs at the radius extrema themselves.
The star is most radially extended near minimum light, when it is coolest, and most contracted near maximum light, when it is hottest: here the maximum-radius timestep has $T_{\rm eff} = 6214$~K at $R = 5.78\,R_\odot$, while the minimum-radius timestep is hotter, $7486$~K at $R = 4.99\,R_\odot$.
That the radius extrema carry a $\sim$1300~K temperature contrast is the phase offset of Figure~\ref{fig:rsp_radius_diagnostic} expressed in the SED.
The residual $\Delta F_\lambda = F_{\rm max\,R} - F_{\rm min\,R}$ is therefore dominated by this temperature difference in the blue, where the cooler maximum-radius state is fainter despite its larger area; toward the infrared the geometric $R^2$ gain of the larger star increasingly offsets its lower temperature, the residual crosses zero near F158, and the reddest bands are marginally brighter at maximum radius.
This combination of a temperature term and a geometric term emerges only because \texttt{MESA Custom Colors} evaluates the full interpolated SED at the actual $T_{\rm eff}$ and $R$ of each timestep, rather than applying a temperature-independent $R^2$ scaling.

\subsubsection[Validation Against Data: Fitting FN Lyr / KIC 6936115]{Validation Against Data: Fitting FN~Lyr / KIC\,6936115}
\label{sec:rsp_fnlyr}

We test against FN~Lyr (KIC\,6936115), a non-Blazhko fundamental-mode RRab star observed by \textit{Kepler} with high precision and a stable light curve \citep{Nemec2011,Li2014}.
Its period is known to nine significant figures, $P = 0.527398471$~d \citep{Nemec2011}, and its \textit{Kepler}-band total amplitude is $A = 1.046$~mag.
\citet{Nemec2011} also derived physical parameters from Fourier decomposition of the same light curve: $T_{\rm eff} \approx 6480$--$6560$~K, $[\mathrm{Fe/H}] \approx -1.6$, and $M = 0.595$--$0.69\,M_\odot$ and $L = 40$--$56\,L_\odot$.
The pulsational and evolutionary estimates disagree at the $\sim$15\% level in both mass and luminosity, a long-standing tension for RR~Lyrae and Cepheids \citep{Nemec2011}.

We fit the relative \textit{Kepler}-band light curve.
The period is a directly measured dynamical observable, and the metallicity comes from the Fourier $\phi_{31}$ relation, which is largely independent of the amplitude and detailed shape we are fitting; we therefore treat the measured period as a fixed fitting target and adopt $Z = 0.0004$ ($[\mathrm{Fe/H}] \approx -1.6$).
By contrast, the published $M$, $L$, and $T_{\rm eff}$ are themselves derived from this light curve through empirical Fourier calibrations, so imposing them as inputs would be circular.
We therefore use these estimates to guide the search rather than imposing them as hard bounds, and judge the fit on three independent observables of the waveform itself: the limit-cycle period, the amplitude, and the detailed light-curve morphology.

The model light curve is the output of a nonlinear RSP integration: its shape is fixed by the physical inputs and cannot be slid or warped, so each candidate is compared to the data after only a phase shift and a vertical offset.
The published estimates guide a coarse grid in $(T_{\rm eff}, L, M)$ and the initial velocity kick, but do not define hard boundaries: the grid spans $T_{\rm eff}=6450$--$6700$~K, while $M$ and $L$ remain within their quoted ranges.
We then carry out targeted sweeps of the RSP convective parameters and automated Nelder--Mead refinements (Appendix~\ref{app:fnlyr_campaign}); the model search comprises roughly fifty settled models in total.
Amplitude is always measured at the settled limit cycle, never during the kick-driven transient, where it tracks the kick rather than the physics.\footnote{We accept a cycle only once its peak-to-peak amplitude is stationary to $\lesssim 10^{-3}$~mag over the final ten cycles.}
Each model is scored on three equally weighted axes: fractional period error, fractional amplitude error, and a morphology metric defined as the RMS residual between the amplitude-normalized model and observed light curves (so that amplitude, scored separately, does not re-enter, and a model is penalized only where its \emph{shape} departs from the real star).

\begin{figure}[tbp]
	\centering
	\includegraphics[width=\linewidth]{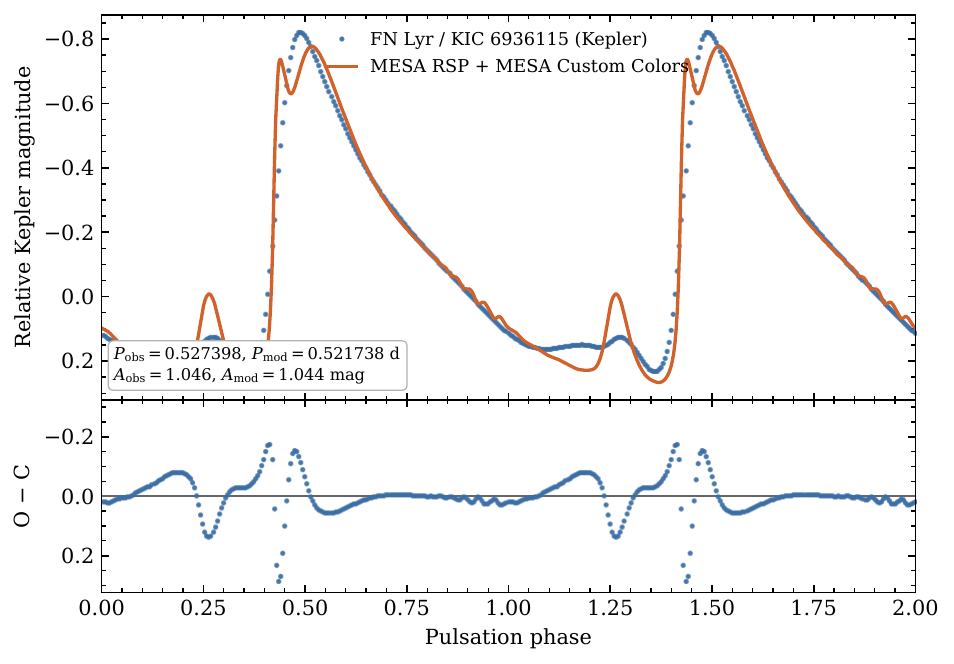}
	\caption{\texttt{MESA} RSP model with \texttt{MESA Custom Colors} fitted to the \textit{Kepler} light curve of FN~Lyr (KIC\,6936115).
		\textit{Top}: observed binned \textit{Kepler} photometry (points) and the best-fit synthetic light curve (line) over two pulsation phases, aligned by phase shift and vertical offset only.
		\textit{Bottom}: residuals (observed $-$ model).
		The model recovers the period to $1.1\%$, the amplitude to $0.2\%$, and the overall RRab morphology; the principal residual is the bump-and-shoulder structure on the rise (see text).}
	\label{fig:fnlyr_fit}
\end{figure}

The best-balanced model across the model search has $T_{\rm eff} = 6700$~K, $L = 47.7\,L_\odot$, $M = 0.65\,M_\odot$, eddy-viscosity coefficient $\alpha_{\rm m} = 0.15$, and turbulent-flux coefficient $\alpha_{\rm t} = 0.01$.
It reproduces the period to $1.1\%$ ($P_{\rm mod} = 0.5217$~d) and the amplitude to $0.2\%$ ($A_{\rm mod} = 1.044$~mag against the observed $1.046$~mag), and it has the lowest combined score across all three axes of any model in the model search.
Figure~\ref{fig:fnlyr_fit} compares the synthetic and observed light curves over two pulsation phases, with residuals; the model reproduces the characteristic RRab waveform with a median absolute residual of $0.02$~mag after phase and offset alignment alone.
The recovered mass and luminosity sit on the \emph{evolutionary} side of the \citet{Nemec2011} values ($M = 0.65\,M_\odot$, $L = 47.7\,L_\odot$), while the effective temperature, $6700$~K, lies somewhat above the Fourier-derived range.

The principal residual, the bump-and-shoulder structure on the rise, originates in the pulsation model rather than in the photometry.
It is set by the RSP eddy-viscosity parameter $\alpha_{\rm m}$: across our model search, matching the observed amplitude required $\alpha_{\rm m} \approx 0.15$, below the value $0.25$ of the fiducial \texttt{MESA}-RSP convective sets, and the lower eddy viscosity is what sharpens the bump while leaving the period almost unchanged.
A targeted sweep of the turbulent-flux coefficient $\alpha_{\rm t}$ at fixed $\alpha_{\rm m} = 0.15$ found that $\alpha_{\rm t} = 0.01$ marginally reduces the amplitude error without degrading the period or shape scores, and this value is adopted in the reported model.
The recovered mass and luminosity correspondingly fall on the evolutionary rather than the pulsational side of the \citet{Nemec2011} values, echoing the pulsational--evolutionary tension those authors report for FN~Lyr; we do not attempt to resolve it here.
The model search and the three-axis scoring used to select the model are summarized in Appendix~\ref{app:fnlyr_campaign}.

\subsection{Starspots: Chromatic Signatures of Surface Inhomogeneities}
\label{sec:demo_starspots}

Magnetically active cool stars carry dark photospheric regions where strong fields suppress convection, with temperature contrasts of order 10--20\% relative to the quiet photosphere.
Because spots are cooler, they emit less flux at all wavelengths, but the suppression is proportionally stronger in the blue, where the Planck function is most temperature-sensitive.
This chromatic signature is both a diagnostic of magnetic activity and a systematic that must be modeled when interpreting cool-dwarf photometry, particularly for exoplanet transits where spot crossings and rotational modulation can mimic or obscure planetary signals.

\texttt{MESA} implements a modified version of the \texttt{YREC SPOTS} formalism of \citet{Somers2015}, which modifies the atmospheric boundary condition for a two-temperature photosphere.
The spot-modified effective temperature is set by the coverage fraction $f_{\rm spot}$ and the contrast $x_{\rm spot} = T_{\rm spot}/T_{\rm phot}$ through a flux-weighted mean entering the Stefan--Boltzmann boundary condition.
This captures the structural feedback of spots on the evolution and supplies the modified $T_{\rm eff}$ that \texttt{MESA Custom Colors} uses, so the boundary modification propagates self-consistently into the synthetic photometry.
In this runtime demonstration, \texttt{MESA Custom Colors} evaluates a single atmosphere SED at that flux-weighted $T_{\rm eff}$; it does not combine separate quiet-photosphere and spot SEDs. The explicit two-component treatment is used only in Section~\ref{sec:demo_yrec}.

We use BT-Settl atmospheres \citep{Allard2014} rather than the Kurucz default: the grid spans 0.3--1.1\,$M_\odot$, placing the lowest-mass tracks at $T_{\rm eff} \approx 3200$~K, with spot coverage depressing the flux-weighted temperature further below the $\sim$3500~K Kurucz floor.
We adopt LSST $ugrizy$ on the AB system at 10~pc (absolute magnitudes), reflecting the section's motivation: LSST will monitor large numbers of M and K dwarfs in six bands simultaneously \citep{LSST}, a regime in which spot-induced variability is a leading signal.

\begin{figure*}[t]
	\centering
	\includegraphics[width=\linewidth]{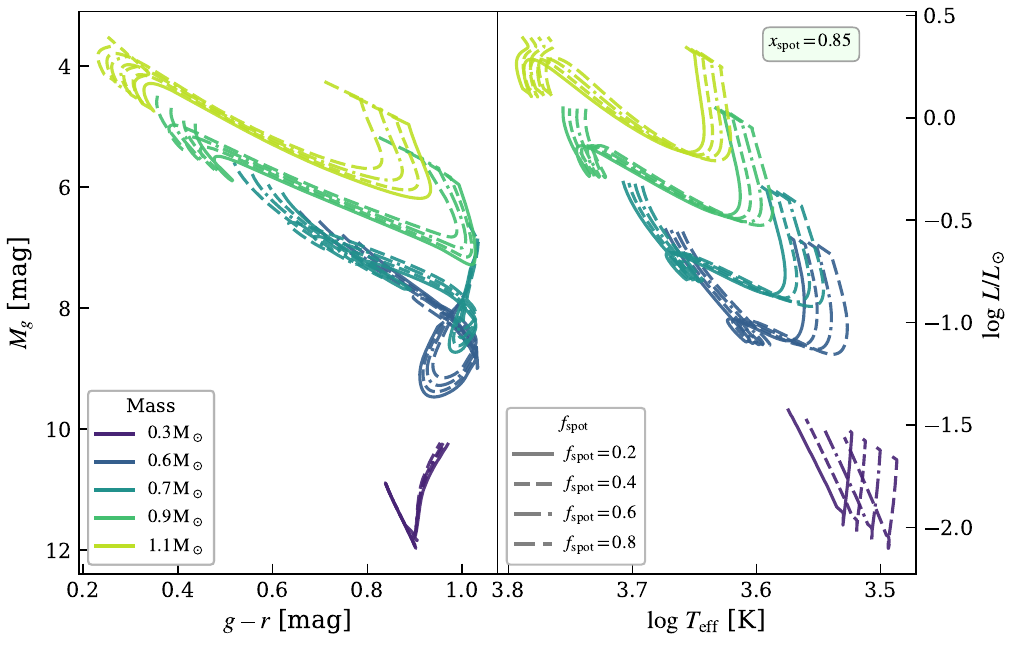}
	\caption{LSST $(g-r)$ versus $M_g$ main-sequence tracks for five masses and four spot filling factors at fixed $x_{\rm spot}=0.85$, the CMD analogue of \citet{Jermyn2023} Figure~15.
		Color encodes mass; linestyle encodes $f_{\rm spot}$.
		Within each mass group, increasing $f_{\rm spot}$ shifts tracks to redder colors and fainter magnitudes.}
	\label{fig:starspots_cmd}
\end{figure*}

Figure~\ref{fig:starspots_cmd} presents the LSST $(g-r)$ versus $M_g$ tracks for the grid of five masses and four spot filling factors ($f_{\rm spot} = 0.2, 0.4, 0.6, 0.8$) at fixed $x_{\rm spot} = 0.85$.
Increasing spot coverage shifts tracks to redder $g-r$ and fainter $M_g$ at fixed mass and age.
The separation between tracks of equal mass but different $f_{\rm spot}$ is larger toward bluer colors: because spot suppression is stronger in $g$ than $r$, the $g$ band brightens relative to $r$ as spots decrease, driving both the $g-r$ reddening and the $M_g$ shift.

This wavelength-dependent displacement provides a multi-band handle on spot properties, although separating it from interstellar reddening or metallicity effects requires modeling those contributions explicitly.
Lower-mass stars, intrinsically cooler, show larger $g-r$ displacements at fixed $f_{\rm spot}$, consistent with the steeper Planck temperature sensitivity at their photospheric temperatures.

The chromatic effect shown here (stronger dimming in blue bands than red) follows from any two-temperature photosphere and is computed in the native survey bands from the same spot parameters used in the evolution.

Section~\ref{sec:demo_yrec} returns to starspots from the complementary angle of rotational modulation, using the stand-alone \texttt{SED\_Model} interface on an external \texttt{YREC} grid.

\subsection{White Dwarfs: Specialized Atmosphere Grids at Extreme Surface Gravity}
\label{sec:demo_wd}

White dwarf cooling sequences are fundamental chronometers for stellar populations, with the luminosity function encoding star-formation history through the progressive fading of remnants over gigayear timescales \citep[e.g.][]{Winget1987,Fontaine2001,Hansen2004}.
Surveys now detect white dwarfs in numbers large enough that comparing cooling tracks to observed color--magnitude distributions is a viable test of cooling and atmosphere physics \citep[e.g.][]{GentileFusillo2021,Gaia2023}.
For synthetic photometry the central challenge here is that the extreme surface gravities of white dwarfs ($\log g \sim 7$--9) place them outside general-purpose atmosphere grids.
The Kurucz/ATLAS9 and BT-Settl models used in other demonstrations terminate near $\log g \approx 5$, so applying them to white dwarfs would require extrapolating by several dex in the parameter that controls atmospheric pressure stratification.

We initialize a $0.6\,M_\odot$ CO white dwarf from a thermally settled model and evolve the cooling sequence with \texttt{use\_colors = .true.} and \texttt{history\_interval = 1}.
Rather than the Kurucz grid of the blue-loop test or the BT-Settl grid of the starspot test, we specify the \citet{Koester2010} pure-hydrogen DA grid via \texttt{stellar\_atm}; it covers $T_{\rm eff} = 5{,}000$--$80{,}000$~K and $\log g = 6.5$--$9.5$, spanning the relevant range at this model's $\log g \approx 7.9$.

As a population-level consistency check, rather than a calibrated fit, we place the cooling model on the observed Gaia white dwarf sequence.
We run the model with the Gaia $G$, $G_{\rm BP}$, and $G_{\rm RP}$ filters on the Vega system, and take the color and absolute magnitude directly from the \texttt{MESA} history columns, with the run at 10~pc so that the resulting $G$ magnitude is absolute.
No empirical color transformation and no external bolometric correction are applied.

The comparison sample is built from Gaia DR3 photometry and astrometry \citep{gaia2016,Gaia2023}.
For each source the absolute magnitude is $M_G = G + 5\log_{10}(\varpi) - 10$ with $\varpi$ the parallax in milliarcseconds.
We require $\varpi > 0$, $\varpi/\sigma_\varpi > 10$, finite $G_{\rm BP}-G_{\rm RP}$, $G < 21$, and $\mathrm{RUWE} < 1.4$ to remove poor astrometric solutions \citep{Lindegren2021}.
The broad CMD sample is supplemented with Gaia white dwarf candidates selected by the Gaia DR3 Discrete Source Classifier, requiring a combined-modules white dwarf probability $\geq 0.5$ \citep[\texttt{classprob\_dsc\_combmod\_whitedwarf};][]{Delchambre2023} under the same astrometric and photometric cuts, so the cooling sequence is well populated while retaining the main-sequence and giant context.
The source density is used only to visualize the observed locus; it is not a volume-complete luminosity function.

\begin{figure*}[t]
	\centering
	\includegraphics[width=\textwidth]{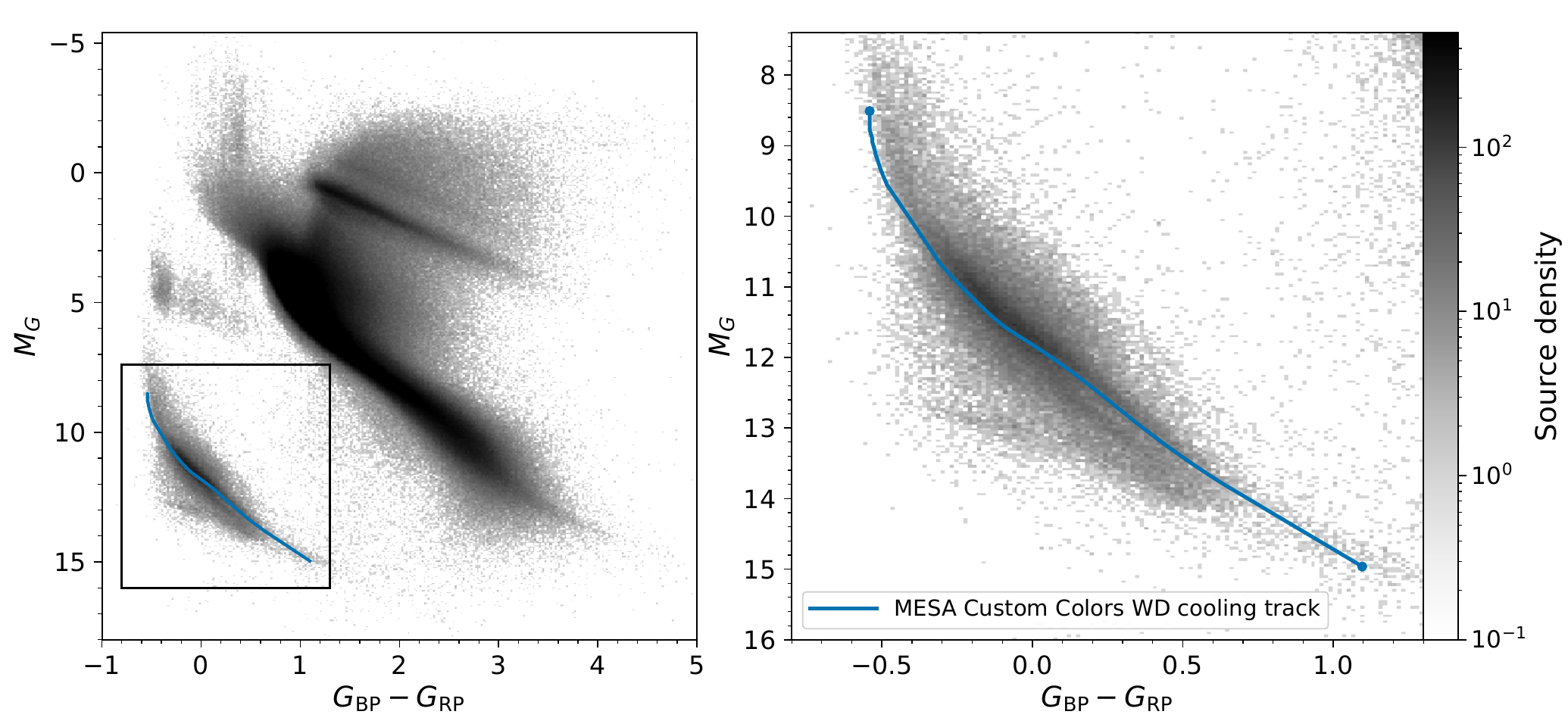}
    \caption{\texttt{MESA Custom Colors} white dwarf cooling sequence against the observed Gaia CMD.
    \textit{Left}: full Gaia $M_G$ versus $G_{\rm BP}-G_{\rm RP}$ diagram, with the model track in cyan; the black rectangle marks the zoom region.
    \textit{Right}: the white dwarf region, where the same track is compared with the observed cooling locus from the hot, blue end to the cool, faint end.
    A binned comparison over $M_G = 9$--15 gives a median absolute color offset of $0.086$~mag, an RMS offset of $0.134$~mag, and a maximum bin-wise offset of $0.273$~mag.
    Model magnitudes are taken directly from the \texttt{MESA} history columns $G$, $G_{\rm BP}$, $G_{\rm RP}$ on the Gaia Vega system, with no post-processing color transformation.
    Source density visualizes the observed locus and is not a volume-complete luminosity function.}
	\label{fig:wd_gaia_cmd}
\end{figure*}

Figure~\ref{fig:wd_gaia_cmd} shows the result.
The track follows the observed cooling locus from the hot blue end ($G_{\rm BP}-G_{\rm RP} \simeq -0.5$, $M_G \simeq 8.5$) to the cool faint end ($G_{\rm BP}-G_{\rm RP} \simeq 1.1$, $M_G \simeq 15$), using the native Gaia filter curves and the \texttt{MESA Custom Colors} output directly.
To quantify this comparison without treating the observed population as a single-mass or volume-complete cooling sequence, we compare the model to the empirical Gaia white dwarf locus in 1-mag bins over $M_G = 9$--15.
In each bin, we compute the median observed $G_{\rm BP}-G_{\rm RP}$ color after clipping outliers, and compare it to the model color interpolated to the median $M_G$ of the Gaia sources in that bin.
Across the six bins, the binned Gaia locus and the $0.6\,M_\odot$ DA \texttt{MESA Custom Colors} track differ by a median absolute color offset of $0.086$~mag, with an RMS offset of $0.134$~mag and a maximum bin-wise offset of $0.273$~mag.
The largest offset occurs in the faintest bin, where the Gaia sample is sparsest and the comparison is most sensitive to the mixture of white dwarf masses, atmosphere compositions, unresolved binaries, and selection effects.

\subsection{Blue Loops: Large-Amplitude Chromatic Evolution}
\label{sec:demo_blueloop}

Intermediate-mass stars (3--9\,$M_\odot$) execute large temperature excursions during core helium burning, with blue loops carrying them from $T_{\rm eff} \sim 4300$~K to $\sim$7700~K and back at nearly constant luminosity.
This $\sim$3400~K swing produces color changes of $\Delta(g-r) \sim 1.0$~mag in this model. More generally, such temperature excursions carry intermediate-mass stars across the classical Cepheid instability strip \citep{Tarczay2026}.
Predicting the multi-band photometry requires atmosphere interpolation across a wide temperature range at the moderate surface gravities ($\log g \approx 1.4$--2.4) of evolved giants, so the blue loop provides a broad visual check of interpolation behavior across the sampled sequence.

Each crossing of the instability strip places the star in the regime of radiatively driven pulsation that produces the period--luminosity relation underpinning the extragalactic distance ladder.
The number of crossings, their duration, and the luminosity at each depend on mass, metallicity, and mixing, so Cepheid population predictions need photometry computed at the exact timesteps the model is within the strip.

We use a $5\,M_\odot$, $Z = 0.008$ model evolved through core helium burning with \texttt{use\_colors = .true.} and \texttt{history\_interval = 1}, with LSST $ugrizy$ and Kurucz atmospheres.

\begin{figure*}[t]
	\centering
	\includegraphics[width=\linewidth]{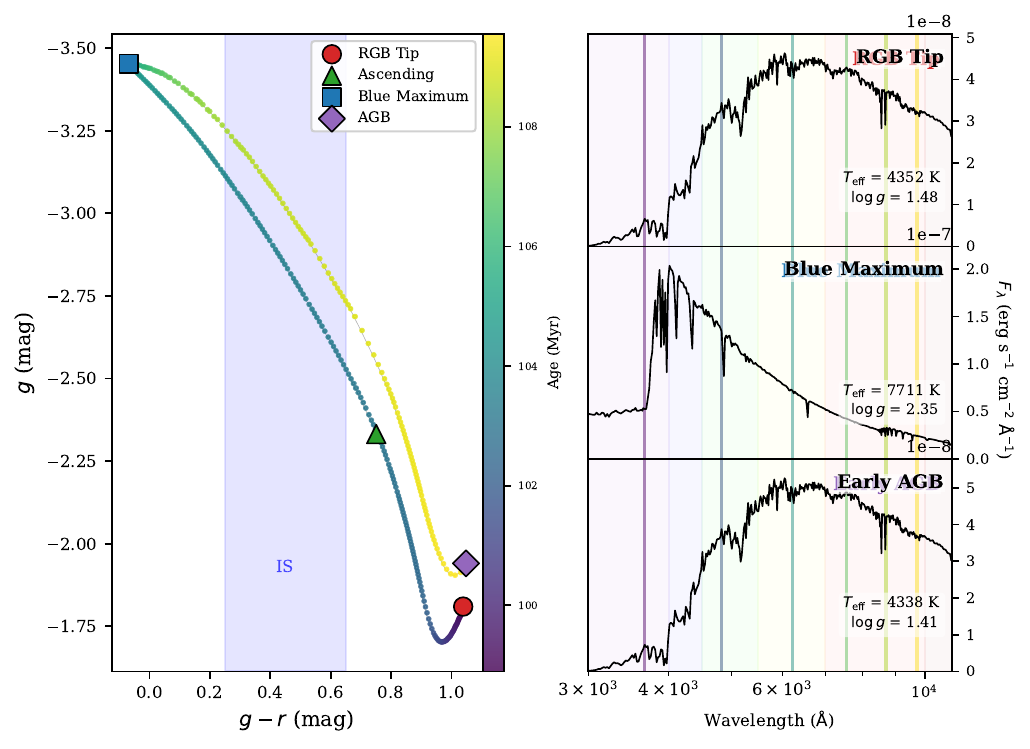}
	\caption{\textit{Left}: blue-loop trajectory in LSST $(g-r)$ versus $g$ for a $5\,M_\odot$, $Z=0.008$ model, colored by age; the shaded region is the instability strip.
		\textit{Right}: SEDs at three marked phases, with LSST bandpasses shaded; the spectral peak shifts from $\sim$7000~\AA\ at the RGB tip to $\sim$4000~\AA\ at blue maximum.}
	\label{fig:blueloop_cmd}
\end{figure*}

Figure~\ref{fig:blueloop_cmd} shows the $(g-r)$ versus $g$ evolution: the star reddens along the RGB, reaches $(g-r) \approx 1.0$ at the tip, makes a rapid blueward excursion during core helium burning, then returns toward the AGB.

\begin{figure*}[t]
	\centering
	\includegraphics[width=\linewidth]{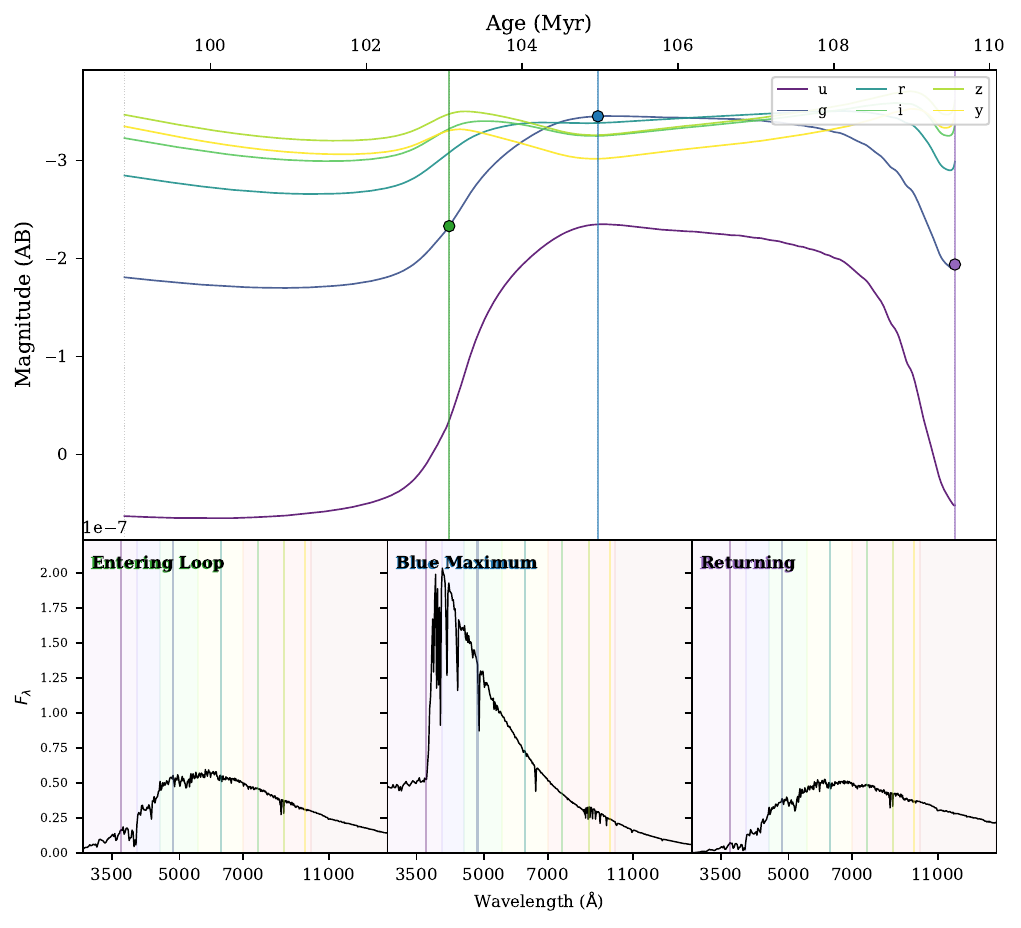}
	\caption{\textit{Top}: multi-band LSST light curves during blue-loop evolution, with vertical lines at three epochs.
		\textit{Bottom}: SEDs at those epochs, from red-peaked (entering) through blue-peaked (maximum) and back (returning).
		Bandpasses are shaded.}
	\label{fig:blueloop_multiband}
\end{figure*}

Figure~\ref{fig:blueloop_multiband} shows the corresponding light curves: the $u$ band varies by $\sim$2.5~mag while the redder bands vary by $\lesssim$0.5~mag, again reflecting the stronger temperature sensitivity of bluer passbands.

\subsection{Starspot Rotational Modulation: Post-Processing the YREC SPOTS Grid}
\label{sec:demo_yrec}

The starspot demonstration of Section~\ref{sec:demo_starspots} used \texttt{MESA}'s runtime boundary condition to produce spot-modified photometry during the evolution.
Here we take the complementary path: \texttt{SED\_Model}, the stand-alone Python interface to the same Fortran routines used by \texttt{MESA Custom Colors}, post-processes the \texttt{YREC} Stellar Parameters of Tracks with Starspots (SPOTS) grid \citep{Somers2020}.
This exercises the custom-physics capability described in Section~\ref{sec:intro}: the identical photometry engine operates on an external evolutionary calculation that \texttt{MESA} never produced, requiring no \texttt{MESA} run at all.

The SPOTS grid \citep{Somers2015,Somers2020} comprises 150 tracks spanning 25 masses (0.10--1.30\,$M_\odot$ in steps of $0.05\,M_\odot$) and six covering fractions ($f_{\rm spot} = 0, 0.17, 0.34, 0.51, 0.68, 0.85$) at fixed contrast $x_{\rm spot} = T_{\rm cool}/T_{\rm hot} = 0.80$, computed with the Yale Rotating Evolution Code \citep{Demarque2008}.
Each track records $T_{\rm hot}$ and $T_{\rm cool}$ separately at every timestep rather than only the flux-weighted blend.
\texttt{SED\_Model} exploits this by running the forward model on each component, retaining both SEDs at each step; the two are then combined at arbitrary spot visibility fraction to predict the rotationally modulated light curve with no further stellar modeling.
We use Kurucz/ATLAS9 atmospheres with Johnson $UBVRIJ$ in the AB system at 10~pc.

\begin{figure*}[t]
	\centering
	\includegraphics[width=\linewidth]{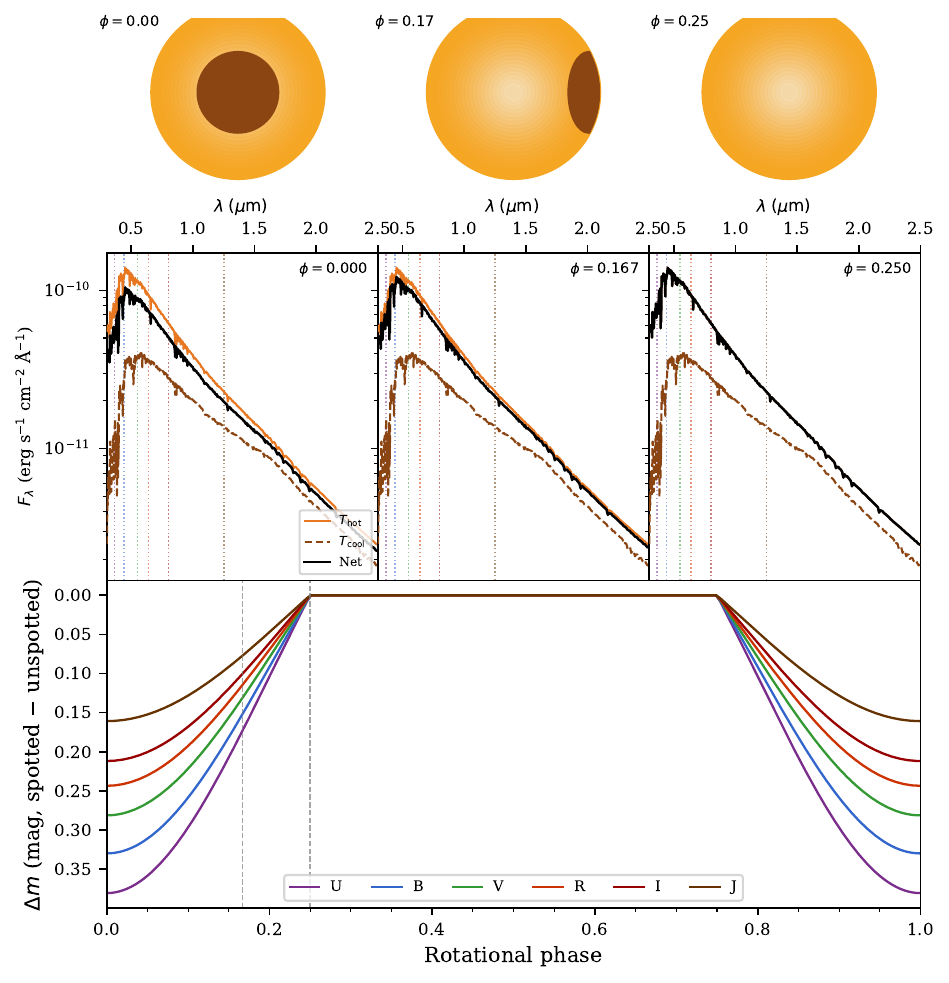}
	\caption{Rotational modulation of a $1.10\,M_\odot$ main-sequence star with $f_{\rm spot} = 0.34$, $x_{\rm spot} = 0.80$.
		\textit{Top}: stellar disc at three rotational phases; the spot moves from disc center ($\phi = 0$) to the limb ($\phi = 0.25$).
		\textit{Middle}: photospheric SED ($T_{\rm hot}$, solid orange), spot SED ($T_{\rm cool}$, dashed brown), and net (black); Johnson pivot wavelengths as vertical dotted lines.
		\textit{Bottom}: Johnson $UBVRIJ$ light curves versus phase.
		The $U$-band amplitude ($\approx 0.37$~mag) exceeds $J$ ($\approx 0.15$~mag) by $\sim$2.5, reflecting the stronger blue temperature sensitivity.}
	\label{fig:yrec_modulation}
\end{figure*}

Figure~\ref{fig:yrec_modulation} illustrates the two-component approach for a $1.10\,M_\odot$ model near the end of the main sequence ($f_{\rm spot} = 0.34$, $x_{\rm spot} = 0.80$).
The top row shows the stellar disc at three rotational phases as the spot moves from disc center ($\phi = 0$) to the limb ($\phi = 0.25$); the middle row shows the component SEDs ($T_{\rm hot}$, $T_{\rm cool}$) and their visibility-weighted combination; the bottom row shows the six-band light curves.
As the spot rotates to the limb, the net SED converges to the photospheric one, faster in the infrared (where the components differ least) than in the blue.
At $\phi = 0$ the $U$-band dimming reaches $\Delta m \approx 0.37$~mag while $J$ reaches only $\approx 0.15$~mag, a factor of $\sim$2.5 set entirely by the wavelength dependence of the spot-to-photosphere contrast.

\begin{figure}[tbp]
	\centering
	\includegraphics[width=\columnwidth]{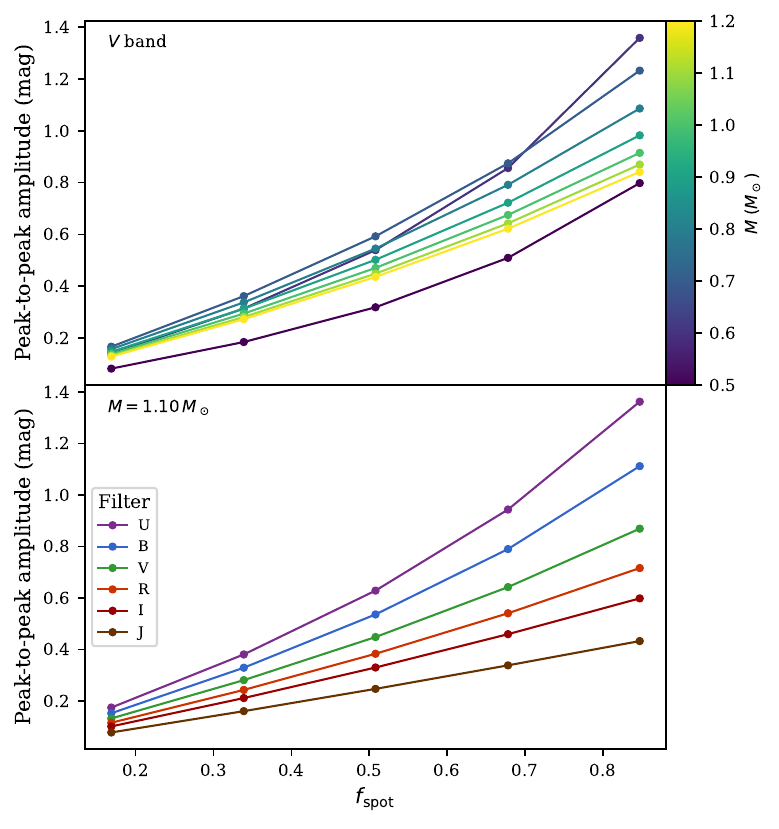}
	\caption{Peak-to-peak rotational modulation amplitude versus spot covering fraction $f_{\rm spot}$ from the YREC SPOTS grid at 3~Gyr.
		\textit{Top}: $V$-band amplitude, colored by mass.
		\textit{Bottom}: all six Johnson filters at $1.10\,M_\odot$; $U$ exceeds $J$ by $\sim$3 at $f_{\rm spot} = 0.85$.}
	\label{fig:yrec_amplitude_grid}
\end{figure}

Figure~\ref{fig:yrec_amplitude_grid} maps this across the grid.
The $V$-band peak-to-peak amplitude rises monotonically with $f_{\rm spot}$, reaching $\sim$1.35~mag at $f_{\rm spot} = 0.85$ for the $1.10\,M_\odot$ model at 3~Gyr.
Higher-mass stars at this age are hotter, placing $V$ further into the Wien regime of both components and raising the $T_{\rm hot}$--$T_{\rm cool}$ contrast at fixed $x_{\rm spot}$.
The lowest-mass tracks ($0.50\,M_\odot$) show reduced amplitudes at fixed $f_{\rm spot}$ because $T_{\rm cool} \approx 3120$~K falls below the Kurucz floor ($\sim$3500~K), clamping the interpolation and suppressing the apparent contrast. A combined Kurucz+BT-Settl grid could potentially recover the full amplitude.
The lower panel separates the six bands at $1.10\,M_\odot$: at $f_{\rm spot} = 0.85$ the $U$ amplitude ($\approx 1.35$~mag) exceeds $J$ ($\approx 0.44$~mag) by a factor of three, with $V$, $R$, $I$ ordered by wavelength in between.

These amplitude surfaces provide forward-model predictions at the parameters monitored by LSST and PLATO \citep{LSST,plato}, where rotational modulation of active dwarfs is a primary systematic.
\citet{Herbert2024} measured peak-to-peak modulation in Johnson $V$, $R$, $I$ for 32 periodic young stellar objects in IC~5070 from the HOYS dataset \citep{Froebrich2018}, fitting a flux-replacement model for spot temperature and covering fraction.
Their cold-spot solutions span $f \approx 0.06$--0.40 with contrasts 280--2150~K below the surface, for hosts at $T_{\rm eff} \approx 3900$--4300~K.
The observed ordering $A_V > A_R > A_I$ is reproduced by our lower panel at all covering fractions, and because \citet{Herbert2024} note that $V$, $R$, $I$ are insensitive to high-temperature accretion shocks and that bluer bands would improve characterization, the $U$ and $B$ predictions here extend their coverage into exactly that regime.

\section{Limitations, Assumptions, and Usage Guidelines}
\label{sec:limitations}

With \texttt{Custom Colors}, synthetic magnitudes are computed from stellar atmosphere grids interpolated in effective temperature, surface gravity, and composition; radiative transfer is not solved.
The fidelity of the resulting photometry is bounded by the coverage, resolution, and internal consistency of the adopted atmosphere models.
Requests outside the grid boundaries are clamped to the nearest edge point and flagged through \texttt{Interp\_rad}; photometry produced under significant clamping should be treated with caution.

Grid interpolation can itself be a limiting source of error, particularly where the emergent spectrum varies rapidly with stellar parameters such as near strong line blanketing, molecular bands, ionization transitions, or other abrupt changes in spectral morphology.
Higher-order schemes produce smoother tracks in color--magnitude space but can introduce non-physical overshoot where local grid behavior is insufficiently smooth. 
Lower-order schemes are more robust but may produce derivative discontinuities as a model crosses grid-cell boundaries.
Synthetic photometry near sharp atmospheric transitions warrants additional scrutiny. 
Bolometric fluxes, colors, and \texttt{Interp\_rad} should vary smoothly across the relevant portion of the evolutionary sequence.

The composition dependence of the synthetic photometry is limited by the abundance dimensions of the selected atmosphere grid.
Although \texttt{MESA} tracks the evolving surface composition, most atmosphere grids represent composition through a single metallicity coordinate with a fixed abundance pattern.
Where an adopted atmosphere grid is scaled-solar, its $\mathrm{[M/H]}$ coordinate may be identified with $\mathrm{[Fe/H]}$; the two are not generally interchangeable for non-solar abundance patterns.
Changes in individual abundance ratios, $\alpha$-enhancement, C/O ratio, surface pollution, or transitions between atmospheric classes are captured only where explicitly represented in the chosen atmosphere library.
This limitation is particularly relevant for chemically peculiar stars, carbon-rich or oxygen-rich AGB stars, very metal-poor populations, and white dwarfs whose surface composition departs from the grid assumptions.
In such regimes, smooth interpolation and small \texttt{Interp\_rad} values are numerical diagnostics only and do not indicate that the adopted atmosphere model is physically appropriate.

Distances are fixed and user-specified.
Binary orbital motion, time-variable distances, and flux contributions from unresolved companions are not modeled\footnote{This task can be implemented by the user through \texttt{run\_star\_extras.f90}.}.
No correction for interstellar extinction or reddening is applied; comparisons to observed magnitudes therefore require external corrections appropriate to the chosen filters and line of sight.
Support for extinction and reddening is planned for a future release, but these values will still need to be provided by the user for each object or application.

Photometric calculations assume idealized filter transmission curves and do not account for time-dependent instrumental effects, detector systematics, or variation in observational throughput.
Synthetic photometry is evaluated only at points where \texttt{MESA} writes history output, so the effective temporal resolution is governed by \texttt{history\_interval}.
Rapidly changing evolutionary or pulsational states may therefore require \texttt{history\_interval = 1} so that photometry is written for every converged timestep.

Formal photometric uncertainties arising from atmosphere-grid choice, filter definitions, zero-point conventions, and interpolation error are not propagated.
Sensitivity tests using alternative grids, filter curves, magnitude systems, or interpolation schemes provide the most direct means of quantifying these contributions.

\section{Summary and Future Work}
\label{sec:conclusions}

At each history-output timestep, \texttt{MESA Custom Colors} converts the current stellar surface state into observer-frame photometry by interpolating an atmosphere grid, applying the requested filter curves, and writing the resulting magnitudes to \texttt{history.data}.
Configuration is through a \texttt{\&colors} namelist block, allowing the atmosphere grid, filter system, photometric zero-point convention, and source distance to be recorded alongside the stellar physics. Reproducing a calculation requires the \texttt{inlist}, the \texttt{MESA} revision, and the same atmosphere grid and filter files, since the namelist records paths to those data rather than their contents.

Six demonstrations exercise the module across TP-AGB evolution, nonlinear RR~Lyrae pulsation, starspots, white dwarf cooling, blue loops, and application of the photometry routines to an external \texttt{YREC SPOTS} grid via the \texttt{SED\_Model} Python package.
The tests span three atmosphere grid families (Kurucz/ATLAS9, BT-Settl, and Koester DA), filter systems including Roman WFI, LSST $ugrizy$, Gaia $G$/$G_{\rm BP}$/$G_{\rm RP}$, and Johnson $BVRI$/$UBVRIJ$, and two photometric systems (AB and Vega); grid and filter substitution requires only namelist edits.

The two real-data comparisons provide the most direct tests. 
Against the \textit{Kepler} light curve of FN~Lyr (KIC\,6936115), a settled RSP model recovers the period to $1.1\%$ and amplitude to $0.2\%$ after phase and vertical alignment. The residual rise morphology is attributed to the RSP eddy-viscosity parameter $\alpha_{\rm m}$ rather than to the photometry (Section~\ref{sec:rsp_fnlyr}, Appendix~\ref{app:fnlyr_campaign}).
A $0.6\,M_\odot$ DA white dwarf cooling track produced from \texttt{MESA} history columns in the native Gaia filter system follows the observed DR3 cooling locus from $M_G \simeq 8.5$ to $15$~mag with no intermediate color transformation.

The $2\,M_\odot$ TP-AGB model was sampled at each flash rise directly from the saved stellar states, and the $V$-band brightening at pulse peak was shown to be radius-driven rather than temperature-driven.
The starspot demonstration recovers stronger flux suppression in blue bands than red in LSST $ugrizy$, producing CMD displacements toward redder colors and fainter magnitudes. 
The Kurucz/ATLAS9-interpolated blue-loop track appears smooth over the sampled $\sim$3400~K excursion, with $u$-band amplitude exceeding the reddest LSST band by more than a factor of five. 
Post-processing the \texttt{YREC SPOTS} grid through the \texttt{SED\_Model} photometry routines yields chromatic amplitude ordering $U > B > V > R > I > J$ at fixed covering fraction, consistent with the Johnson $VRI$ ordering measured by \citet{Herbert2024} for young spotted stars in IC~5070.

\texttt{SED\_Tools} is the companion Python package for constructing atmosphere grids in the format the module consumes.
It downloads model spectra from public catalogs (the Spanish Virtual Observatory, MSG grids, and MAST/BOSZ) standardizes them onto common physical units, and packages the result into the \texttt{lookup\_table.csv} and \texttt{flux\_cube.bin} products consumed at runtime.
\texttt{SED\_Tools} also acquires photometric filter transmission curves from the SVO Filter Profile Service \citep{Rodrigo2020} in the format the module expects. The \texttt{SED\_Model} Python package separately exposes the same Fortran routines via \texttt{f2py} for offline use on tracks from any source.

The principal planned extensions are built-in interstellar extinction with multiple reddening laws, additional abundance axes ($\alpha$-enhancement and non-solar C/O) beyond the single metallicity coordinate of current grids, binary flux contributions, and direct coupling to GYRE so that mode frequencies can be evaluated alongside the synthetic photometry.
Predicting photometric pulsation amplitudes and phase lags would require additional radiative-transfer treatment beyond either RSP or GYRE.
The FN~Lyr Nelder--Mead search (Appendix~\ref{app:fnlyr_campaign}) demonstrates that \texttt{MESA Custom Colors} photometry is a well-behaved objective function for an external optimizer: repeated \texttt{MESA} RSP evaluations, scored against the observed light curve, converge on well-fitting parameter combinations with no modification to \texttt{MESA}'s internal solver. A natural extension is to formalize this approach into a forward-modeling pipeline that iterates \texttt{MESA} models against observational data more broadly, following the same strategy applied here to FN~Lyr.

\software{\texttt{MESA} \citep{Paxton2011,Paxton2013,Paxton2015,Paxton2018,Paxton2019,Jermyn2023}, RSP \citep{Smolec2008}, \texttt{SED\_Tools}, \texttt{SED\_Model}, NumPy, SciPy, Matplotlib.}

\section*{Acknowledgments}

This work uses \texttt{MESA} version \texttt{r26.04.1} \citep{Paxton2011,Paxton2013,Paxton2015,Paxton2018,Paxton2019,Jermyn2023}, compiled with \texttt{MESA} SDK version \texttt{26.3.2} \citep{MESASDK}.
The \texttt{MESA} equation of state is a blend of the OPAL \citep{Rogers2002}, SCVH \citep{Saumon1995}, FreeEOS \citep{Irwin2004}, HELM \citep{Timmes2000}, and PC \citep{Potekhin2010} equations of state.
Radiative opacities are primarily from OPAL \citep{IglesiasRogers1996, Rogers2002}, with low-temperature data from \citet{Ferguson2005} and the high-temperature, Compton-scattering-dominated regime from \citet{Buchler1976}. Electron conduction opacities are from \citet{Cassisi2007}.
Nuclear reaction rates are from JINA REACLIB \citep{Cyburt2010}, plus additional tabulated weak reaction rates from \citet{Fuller1985}, \citet{Oda1994}, and \citet{Langanke2000}. Screening is included via the prescription of \citet{Chugunov2007}.
Thermal neutrino loss rates are from \citet{Itoh1996}.
The RSP module used in Section~\ref{sec:rsp_fnlyr} is described in \citet{Smolec2008}.
This research has made use of the Spanish Virtual Observatory (\url{https://svo.cab.inta-csic.es}) project funded by MCIN/AEI/10.13039/501100011033 through grant PID2023-146210NB-I00.
We thank the \texttt{MESA} Team for the development and maintenance of \texttt{MESA}.

\clearpage
\appendix
\twocolumngrid

\section{Data Availability}
\label{sec:data_availability}

\texttt{MESA Custom Colors} is distributed as part of the public \texttt{MESA} source release.
The \texttt{SED\_Tools} and \texttt{SED\_Model} packages used to prepare atmosphere grids, filter sets, and stand-alone photometry for this work are publicly available at \url{https://github.com/nialljmiller/SED_Tools} and \url{https://github.com/nialljmiller/SED_Model}, respectively.
The Gaia DR3 photometry and astrometry used in Section~\ref{sec:demo_wd} are publicly available from the Gaia Archive (\url{https://gea.esac.esa.int/archive/}).
The \textit{Kepler} photometry of FN~Lyr (KIC\,6936115) used in Section~\ref{sec:rsp_fnlyr} is publicly available from the Mikulski Archive for Space Telescopes (MAST).
The \texttt{MESA} \texttt{inlist} files, atmosphere-grid configurations, and analysis scripts used to produce the demonstrations in this paper are available from the corresponding author upon request and will be archived on Zenodo upon publication.

\section{Example \texttt{inlist}}
\label{app:config}

When enabled, \texttt{MESA Custom Colors} computes synthetic photometry from the evolving stellar model and appends the resulting magnitudes and colors as extra columns in \texttt{history.data}.
The only required control is the master switch \texttt{use\_colors = .true.}; the remaining inputs select a filter set, an atmosphere grid, the magnitude system, the source distance, and (optionally) diagnostic outputs.
A complete listing of the namelist parameters and their defaults is given in \texttt{colors/defaults/colors.defaults}; the controls most users will set are:

\begin{itemize}
	\item \texttt{use\_colors} --- master switch; \texttt{.true.} enables all bolometric and synthetic-photometry output.
	\item \texttt{instrument} --- path to a filter-set directory, structured as
	      \texttt{facility/instrument} (e.g.\
	      \texttt{data/colors\_data/filters/Generic/Johnson}). Each filter in the
	      directory becomes a history column.
	\item \texttt{stellar\_atm} --- path to a packaged atmosphere-grid directory
	      (e.g.\ \path{data/colors_data/stellar_models/Kurucz2003all/}). Must
	      contain \texttt{lookup\_table.csv}, the SED files, and optionally
	      \texttt{flux\_cube.bin}.
	\item \texttt{distance} --- source distance in centimeters; the default
	      $3.0857\times10^{19}$\,cm (10\,pc) yields absolute magnitudes, and
	      setting it to the source distance yields apparent magnitudes.
	\item \texttt{mag\_system} --- photometric zero-point convention, one of
	      \texttt{'Vega'}, \texttt{'AB'}, or \texttt{'ST'}.
	\item \texttt{vega\_sed} --- path to the Vega reference spectrum, required
	      when \texttt{mag\_system = 'Vega'}.
	\item \texttt{z\_over\_x\_ref} --- reference metal-to-hydrogen ratio used to
	      map the photospheric composition onto the atmosphere-grid metallicity
	      axis, $\mathrm{[M/H]} = \log_{10}\!\left[(Z/X)/\texttt{z\_over\_x\_ref}\right]$.
	      The default, $2.30057\times10^{-2}$, matches the GS98 solar mixture of
	      the default \texttt{Kurucz2003all} grid.
\end{itemize}

For inspection and debugging, \texttt{make\_csv = .true.} writes the full SED and the filter-convolved fluxes to \texttt{colors\_results\_directory}; with
\texttt{sed\_per\_model = .true.} the model number is appended to each filename so the entire SED history is retained rather than overwritten (this can produce
a very large number of files).
Photometry is exposed through the standard history-column mechanism, so the sampling cadence in \texttt{history.data} is set by the run's history-output controls, principally \texttt{history\_interval}.

\subsection{Working Example}
\label{app:config_excerpt}

The following \texttt{\&colors} block is the excerpt from the
\texttt{inlist\_colors} example distributed with the \texttt{MESA Custom Colors} test suite.
The setup enables \texttt{MESA Custom Colors}, sets the filter system to `Johnson'
(paths are given relative to the \texttt{MESA} installation's \texttt{data/colors\_data}
directory), and sets the stellar atmosphere to `Kurucz2003', the distance
to $10$\,pc, and an AB zero-point system.
CSV output is enabled, but \texttt{sed\_per\_model} is left at its default
of \texttt{.false.}, so the module overwrites the same CSV file (one
per filter) on each call rather than appending the model number to the
filename.

\lstdefinestyle{mesainlist}{
	basicstyle=\ttfamily\scriptsize,
	breaklines=true,
	breakatwhitespace=false,
	columns=fullflexible,
	keepspaces=true,
	xleftmargin=2pt,
	xrightmargin=2pt,
	aboveskip=0.6\baselineskip,
	belowskip=0.6\baselineskip
}

\begin{lstlisting}[style=mesainlist]
&colors
  ! Master switch.
  use_colors = .true.

  ! Filter directory: facility/instrument, with an index file
  ! and one .dat transmission curve per filter.
  instrument = 'data/colors_data/filters/Generic/Johnson'

  ! Atmosphere grid: contains lookup_table.csv, SED files,
  ! and optionally flux_cube.bin for fast interpolation.
  stellar_atm = 'data/colors_data/stellar_models/Kurucz2003all/'

  ! Source distance in cm. 10 pc gives absolute magnitudes.
  distance = 3.0857d19

  ! Zero-point system: 'Vega', 'AB', or 'ST'.
  mag_system = 'AB'

  ! Required only when mag_system = 'Vega'.
  vega_sed = 'data/colors_data/stellar_models/vega_flam.csv'

  ! Optional diagnostic SED and filter-flux output.
  make_csv = .true.
  colors_results_directory = 'SED'
  sed_per_model = .false.
/
\end{lstlisting}

\section{Filter File Format}
\label{app:filter_format}

A filter set is specified by the \texttt{instrument} directory, structured as
\texttt{facility/instrument}.
The directory contains an index file whose name matches the instrument
basename (e.g.\ \texttt{JWST/MIRI/MIRI}); the index lists one filter filename
per line, and each listed file is read as a transmission curve.

Each transmission file is a two-column table of wavelength and dimensionless
throughput, with a single header line that the reader discards before parsing
the numeric rows. The reader uses list-directed input, so both
whitespace-separated and comma-separated values are accepted; the shipped
\texttt{JWST/MIRI/F1000W.dat} file, for example, has the header
\texttt{Wavelength,Transmission} followed by comma-separated pairs:

\begin{verbatim}
Wavelength,Transmission
87430.0,0.0004
87490.0,0.0012
87550.0,0.0020
\end{verbatim}

Wavelengths must be on a monotonically increasing grid, in the same units as
the atmosphere spectra (Appendix~\ref{app:atm_grid_format}), so that the
filter convolution is physically meaningful;
throughput values are dimensionless and typically lie in the range $0$--$1$.
Each filter becomes a history column named after the filter filename with its
extension removed (more precisely, the portion of the filename before the first
period).

\section{Atmosphere Grid Format}
\label{app:atm_grid_format}

A packaged atmosphere grid is specified by the \texttt{stellar\_atm} directory.
The directory must contain a lookup table, \texttt{lookup\_table.csv}, that maps
each grid point to its spectrum file, and the spectrum files themselves; it may
optionally contain a precomputed \texttt{flux\_cube.bin} for fast interpolation.

\subsection{\texttt{lookup\_table.csv}}
\label{app:lookup_table}

The lookup table is comma-separated text with a header row. The reader requires
the columns
\begin{itemize}
	\item \texttt{file\_name} --- the spectrum filename for that grid point,
	      relative to the atmosphere directory;
	\item \texttt{teff} --- effective temperature in K;
	\item \texttt{logg} --- $\log g$ in cgs;
	\item \texttt{metallicity} --- $\mathrm{[M/H]}$.
\end{itemize}
Additional columns may be present and are ignored. The shipped
\texttt{Kurucz2003all\_\_alpha\_04} table, for instance, also records
provenance and model metadata (\texttt{alpha}, \texttt{lh}, \texttt{vtur},
\texttt{source}, and unit-standardization fields), but only \texttt{file\_name},
\texttt{teff}, \texttt{logg}, and \texttt{metallicity} drive the interpolation.

\subsection{Spectrum File Format}
\label{app:spectrum_format}

Each referenced spectrum file is plain text. Lines beginning with \texttt{\#}
are treated as comments and may carry metadata (the shipped Kurucz files record
$T_{\rm eff}$, $\log g$, metallicity, units, and convective parameters this way).
The reader then parses the remaining lines as two numeric columns:
\begin{enumerate}
	\item wavelength;
	\item surface flux density per unit wavelength.
\end{enumerate}
Values are read with list-directed input. The shipped Kurucz files
(e.g.\ \texttt{Kurucz2003all\_fid19993.txt}) tabulate wavelength in Angstroms
and flux in $\mathrm{erg\,cm^{-2}\,s^{-1}}\,\text{\AA}^{-1}$. The atmosphere wavelength
units must match the filter wavelength units (Appendix~\ref{app:filter_format})
so that the convolution is consistent.

The Vega reference spectrum (\texttt{vega\_sed}) follows a related but distinct
convention: it is comma-separated with a header row and columns
\texttt{WAVELENGTH}, \texttt{FLUX}, and \texttt{CONTINUUM}; the module uses the
wavelength and flux columns and ignores the continuum column. The shipped
\texttt{vega\_flam.csv} gives wavelength in Angstroms and flux in
$\mathrm{erg\,cm^{-2}\,s^{-1}}\,\text{\AA}^{-1}$.

\subsection{Optional Precomputed Cube: \texttt{flux\_cube.bin}}
\label{app:flux_cube}

For fast interpolation the atmosphere directory may include
\texttt{flux\_cube.bin}, a Fortran unformatted stream binary with the layout:
\begin{itemize}
	\item four integers: \texttt{n\_teff}, \texttt{n\_logg}, \texttt{n\_meta},
	      \texttt{n\_lambda};
	\item \texttt{teff\_grid(n\_teff)};
	\item \texttt{logg\_grid(n\_logg)};
	\item \texttt{meta\_grid(n\_meta)};
	\item \texttt{wavelengths(n\_lambda)};
	\item \texttt{flux\_cube(n\_teff, n\_logg, n\_meta, n\_lambda)}.
\end{itemize}
The three axis arrays define the interpolation domain over which an SED is
constructed at an arbitrary $(T_{\rm eff}, \log g, \mathrm{[M/H]})$ within the
supported parameter volume. If a model's photospheric metallicity falls outside
the tabulated range, the module clamps to the nearest available metallicity; if
the hydrogen or metal mass fraction is non-positive, so that
$\log_{10}\!\left[(Z/X)/\texttt{z\_over\_x\_ref}\right]$ cannot be formed, it
falls back to the lowest tabulated metallicity. For the shipped
\texttt{Kurucz2003all} table this range is $\mathrm{[M/H]} = -2.5$ to $+0.5$,
corresponding to a lower edge of $Z/X \approx 7.275\times10^{-5}$ at the default
\texttt{z\_over\_x\_ref}. The wavelength grid in the cube must be consistent with
the units used by the filter curves.

\section{FN~Lyr Fitting}
\label{app:fnlyr_campaign}

The best-fit model quoted in Section~\ref{sec:rsp_fnlyr} is selected from a model search of roughly fifty settled RSP models spanning the physical and convective parameter ranges described below.
This appendix documents the model search and the three-axis scoring used to choose among the models, and makes explicit the trade-off that prevents any single model from minimizing all three observables at once.

Each model is integrated to a settled limit cycle and scored on three equally weighted axes against the binned \textit{Kepler} light curve: fractional period error, fractional amplitude error, and a morphology metric equal to the RMS residual between the amplitude-normalized model and observed curves.
Normalizing both curves to unit amplitude before differencing is essential: it prevents the amplitude from re-entering the morphology score, and it ensures that low-amplitude models, which would otherwise appear to have small absolute residuals, are correctly penalized for having the wrong shape.
The morphology residual is additionally split into a rise contribution and a descending-branch contribution, which localizes the misfit: across the model search the descending branch is reproduced well by all reasonable models, and essentially all of the morphology penalty for the better models lives on the rise, consistent with the eddy-viscosity-driven bump discussed in Section~\ref{sec:rsp_fnlyr}.

The model search proceeded in three stages.
The first was a coarse grid in $(T_{\rm eff}, L, M)$ and initial velocity kick that located the basin matching FN~Lyr's period and amplitude; it sampled $T_{\rm eff}=6450$--$6700$~K, while $M$ and $L$ remained within the ranges quoted by \citet{Nemec2011}.
The second swept the RSP convective parameters $\alpha_{\rm m}$, $\alpha_{\rm t}$, and $\gamma_{\rm r}$ individually at the $6700$~K grid basin: $\alpha_{\rm m}$ controls amplitude and is the dominant lever for light-curve shape; $\alpha_{\rm t}$ (turbulent flux) produced a marginal improvement in amplitude error at $\alpha_{\rm t} = 0.01$; $\gamma_{\rm r}$ (radiative damping) amplified the shock and was rejected.
The third stage comprised an initial automated Nelder--Mead search minimizing the waveform RMS with $T_{\rm eff}$, $L$, $M$, and $\alpha_{\rm m}$ free, followed by two morphology-targeted searches that explored the artificial viscosity coefficient $\alpha_{\rm cq}$ and shock onset threshold $\alpha_{\rm zsh}$ successively.
The recorded evaluations spanned $T_{\rm eff}=6543.9$--$6570$~K in the initial simplex, $6600$--$6850$~K in the $\alpha_{\rm cq}$ search, and $6650$--$6800$~K in the $\alpha_{\rm zsh}$ search.
The two morphology parameters were never varied simultaneously: $\alpha_{\rm cq}$ was fixed at $4$ during the $\alpha_{\rm zsh}$ search.
None of the simplex searches relieved the fundamental trade-off: improving the shape residual required drifting the amplitude away from the observed value.

\begin{figure}[tbp]
	\centering
	\includegraphics[width=\columnwidth]{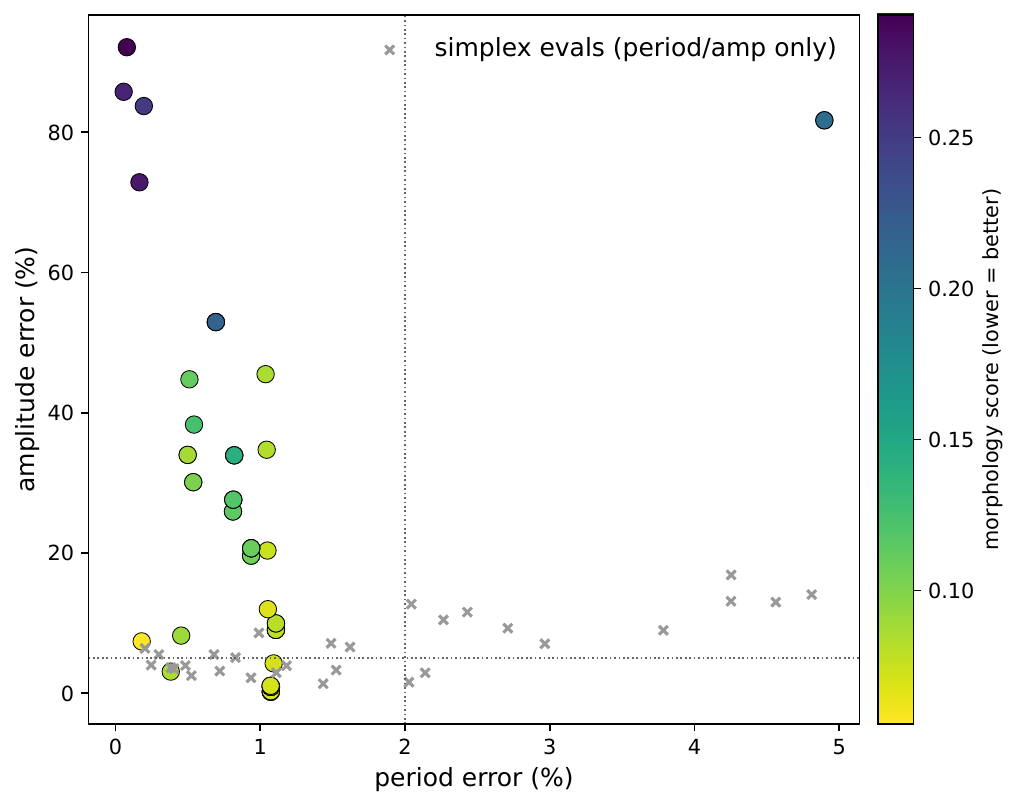}
	\caption{The FN~Lyr model search in the period-error--amplitude-error plane, each model colored by its morphology score (RMS residual of the amplitude-normalized light curve; lower is better).
		Dotted lines mark the $2\%$ period and $5\%$ amplitude levels.
		No model simultaneously minimizes all three axes: the best period/amplitude models carry the largest morphology residuals, reflecting the eddy-viscosity--amplitude coupling discussed in Section~\ref{sec:rsp_fnlyr}.
		Small grey crosses are individual Nelder--Mead evaluations, scored on period and amplitude only (their light curves were not retained for morphology scoring).}
	\label{fig:fnlyr_tradeoff}
\end{figure}

Figure~\ref{fig:fnlyr_tradeoff} shows the model search in the plane of period error versus amplitude error, with each model colored by its morphology score.
The trade-off is explicit: the models that minimize period and amplitude error carry the largest morphology residuals (they sit at the low $\alpha_{\rm m}$, high surface-velocity corner needed for full amplitude), while the smoother-rise models fall short in amplitude.
The reported model ($\alpha_{\rm m} = 0.15$, $\alpha_{\rm t} = 0.01$) is the point that balances all three axes; it is not the unique minimum of any single one.

As an independent check that the synthetic photometry can serve directly as an optimization objective, we use the Nelder--Mead searches described above.
The published physical estimates guided their initialization but were not imposed as hard bounds, and the morphology parameters $\alpha_{\rm cq}$ and $\alpha_{\rm zsh}$ were explored in separate successive searches.
The recorded simplex evaluations reach period errors below $0.5\%$ within a few tens of evaluations, confirming that the \texttt{MESA Custom Colors} light curve is a well-behaved objective function for fitting observed photometry.
None relieves the trade-off of Figure~\ref{fig:fnlyr_tradeoff}: the best morphology-targeted simplex achieved a lower shape residual ($0.056$ versus $0.068$ for the reported model) but at the cost of a $7\%$ amplitude error, as the optimizer drifted $T_{\rm eff}$ and $L$ upward to improve shape while abandoning amplitude accuracy.
It is for this reason that we report the balanced model selected from the full model search rather than the simplex minimum of the waveform residual alone.

\section{Computational Overhead}
\label{app:timing}
\texttt{MESA Custom Colors} adds photometric calculations at each history-output
timestep. To quantify the associated wall-clock cost, we benchmark six
configurations against a module-disabled baseline using a $7\,M_\odot$
pre-main-sequence to ZAMS model evolved for 300 converged timesteps on a
single workstation core. Each configuration was run three times, and the
reported per-step cost is the mean over those repeats. Six configurations are
tested: a \texttt{MESA}-only baseline (\texttt{use\_colors = .false.}); the module
active under each of the three magnitude systems (Vega, AB, ST) without
diagnostic CSV output; the Vega system with \texttt{make\_csv = .true.}; and the
same with \texttt{sed\_per\_model = .true.} additionally enabled.
Table~\ref{tab:timing} summarizes the results.
The benchmark ran on an AMD Ryzen 7 PRO 7840U CPU under Fedora 42, with
\texttt{MESA} r26.04.1 compiled using \texttt{MESA} SDK version 26.3.2.
Wall-clock time was measured around the full run invocation and divided by the
number of converged timesteps, so each figure carries a share of the one-time
initialization cost, including the atmosphere flux-cube load. Over 300 steps
that share is larger than it would be in a longer run.

\begin{table*}[t]
\centering
\caption{Wall-clock cost of each \texttt{MESA Custom Colors} configuration relative to the
         module-disabled baseline, measured over 300 converged timesteps and averaged
         over three repeats. The min--max column gives the range of per-step cost across
         those repeats.
         \label{tab:timing}}
\begin{tabular}{lcccc}
\hline\hline
Configuration & s\,step$^{-1}$ & s\,step$^{-1}$ (min--max) & $\Delta$ s\,step$^{-1}$ & Overhead \\
\hline
Module disabled                & 0.238 & 0.2327--0.2484 & \nodata  & \nodata     \\
Vega, no CSV                   & 0.240 & 0.2387--0.2407 & $+$0.002 & $<$1\%      \\
AB, no CSV                     & 0.239 & 0.2390--0.2392 & $+$0.001 & $<$1\%      \\
ST, no CSV                     & 0.240 & 0.2380--0.2442 & $+$0.002 & $<$1\%      \\
Vega + CSV                     & 0.281 & 0.2791--0.2828 & $+$0.043 & $\sim$18\%  \\
Vega + CSV + per model         & 0.277 & 0.2755--0.2816 & $+$0.039 & $\sim$16\%  \\
\hline
\end{tabular}
\end{table*}

The three configurations without diagnostic CSV output differed from the
module-disabled baseline by at most $0.002$~s per step ($<1\%$). These
differences are smaller than the run-to-run scatter of the baseline itself,
which spanned $0.233$--$0.248$~s per step across its three repeats. The no-CSV
figures are therefore upper bounds rather than resolved measurements, and the
ordering of the three magnitude systems in Table~\ref{tab:timing} carries no
significance.

The largest measured cost was associated with diagnostic CSV output. Enabling
\texttt{make\_csv = .true.} added $0.043$~s per step, about 18\% of the baseline.
Adding \texttt{sed\_per\_model = .true.} gave $0.039$~s per step. The two CSV
configurations differ by less than the baseline scatter and are not
distinguished by this benchmark.

\begin{figure}[tbp]
    \centering
    \includegraphics[width=\columnwidth]{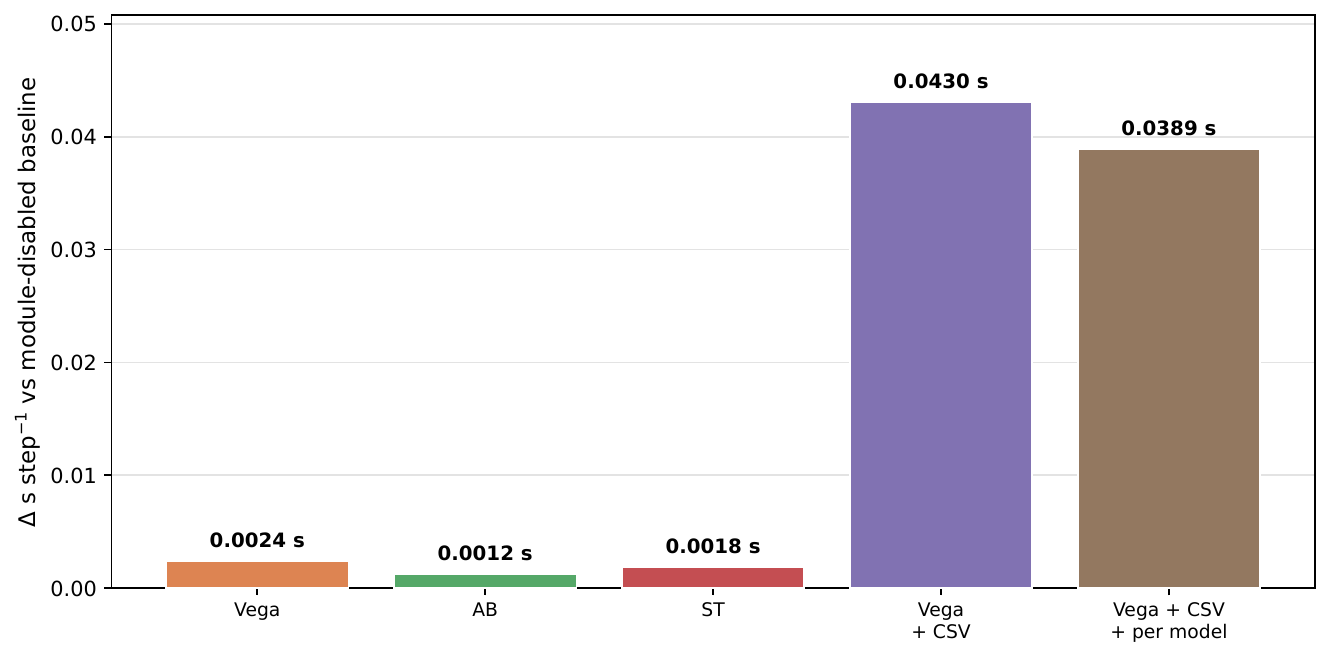}
    \caption{Incremental wall-clock cost of individual \texttt{MESA Custom Colors} features
             relative to the module-disabled (\texttt{use\_colors = .false.}) baseline,
             measured over 300 timesteps of a $7\,M_\odot$ pre-MS model and averaged over
             three repeats. The configurations without CSV output fall below the
             run-to-run scatter of the baseline, whereas enabling \texttt{make\_csv}
             produced the largest increase.}
    \label{fig:timing_breakdown}
\end{figure}

The fractional overheads depend on the cost of the host model's timesteps and
not on the module alone. The model used here averages $0.238$~s per converged
step over the sampled range, so the same absolute CSV cost corresponds to a
smaller fractional overhead in a more expensive calculation. The
$\Delta$~s\,step$^{-1}$ column is the more transferable quantity. These values
characterize this single benchmark setup and should not be interpreted as
platform-independent performance estimates.
\texttt{make\_csv} should therefore be disabled in production runs and reserved
for debugging and spectral inspection.
Figure~\ref{fig:timing_breakdown} shows the per-step cost of each
configuration relative to the module-disabled baseline.

\clearpage
\bibliography{paper}{}
\bibliographystyle{aasjournalv7}

\end{document}